\documentstyle{article}
\def\half{{{1}\over{2}}\; }
\def\rf#1{(\ref{#1})}
\def\L{{\cal L}}
\def\eqn#1#2{\begin{equation} #2
\label{#1}
\end{equation}}
\newtheorem{fignum}[figure]{Fig.}
\def\fref#1{fig.\ref{#1}}
\def\bfref#1{(fig.\ref{#1})}
\def\fig#1#2#3{\begin{fignum}\label{#1}{\rm #3 }\end{fignum} }
\def\nn{\nonumber}
\def\ph{\varphi}

\begin{document}
\title{Singularities on the World Sheets\\ of Open
Relativistic Strings}
\author{S.V.Klimenko, I.N.Nikitin, V.V.Talanov}
\date{}
\maketitle
%\insert\footins{\ \ {\footnotesize *\hspace{0.4cm}E-mail:\ \
%nikitin\_{$\,$}i@mx.ihep.su }}
%\insert\footins{\ \ {\footnotesize **  E-mail:\ \ talanov@mx.ihep.su }}

\begin{abstract}
A classification of stable singular points on world sheets of open relativistic
strings is carried out.
\end{abstract}

\section*{1. Introduction}

String theory considers surfaces in $d$-dimensional Minkowski space, which
have an extreme area -- world sheets of string.
Analogous theory in Euclidean space studies soap films.
String theory considers the world sheets of various topological types:
open strings (obtained by a mapping of a band into Minkowski
space), closed strings (mapping of a cylinder), Y-shaped strings
(mapping of 3 bands, glued together along one edge), etc. A finer
classification takes into account the smoothness of these mappings. It is
known that world sheets can have stable singular points.
An important role of
singularities in string theory was mentioned by many authors.
\begin{itemize}
\item
Singularities have a form of angular points on string.
Density of string's energy-momentum tends to infinity in a singular point.
Near singular points string models of hadrons go beyond the limits of
their applicability. String breaking may occur in singular points \cite{Artru}.
\item
Singularities on the world sheets in 3D space behave like point particles,
that move at the velocity of light, scatter in collisions and can form
bound states. In $\sigma$-model approach to string theory the
singularities manifest themselves as singular solutions of Liouville equation,
they form a completely integrable Hamiltonian system \cite{ECHAQ,smodel}.
Physical interpretation of singularities on string as the elements of
exotic hadrons structure is very attractive, for example, as valent
gluons \cite{Pronkopc}. This identification allows one
to describe exotic hadrons
states in the frames of string model.
\item
Singularities conserve in deformations of the world sheet. In
approaches to string quantization at non-critical dimension \cite{zone}
their number plays a role of super-conserving value (topological charge),
dividing the phase space
into disconnected sectors. For the sector with zero topological charge
the self-consistent quantum theory can be constructed.
\end{itemize}

As far as we know, a detailed classification of singular points on the
world sheets of strings has not been developed yet. The main goal of this work
is a comprehensive study of singularities on the world sheets of open strings.
We have developed a set of visualization tools \cite{zero} for the complicated
three-dimensional surfaces drawing
and used it in this study. We have confirmed the results of earlier
investigations of singularities on world sheets \cite{ECHAQ,Zhelt} and also
observed new phenomena.

\section*{2. World sheet of open string}

Let a periodical isotropic curve
$Q_{\mu}(\sigma)$ be given
in $d$-dimensional Minkowski space:
$$Q'^{2}(\sigma)=0, \qquad Q_{\mu}(\sigma +2\pi)=
Q_{\mu}(\sigma)+2P_{\mu},$$
$2P_{\mu}$ is a period of the curve. This curve is shown on \fref{straight}.
We will refer to this curve as the supporting curve.

The world sheet of open string is constructed as geometrical place of middles
of
segments, connecting all possible pairs of points on supporting curve
\cite{axi}
\begin{equation}
x_{\mu}(\sigma_{1},\sigma_{2})=\half(Q_{\mu}(\sigma_{1})+Q_{\mu}(\sigma_{2})).
\label{sheet}
\end{equation}
Let's prove this. The action of string is proportional to an area of the
world sheet
$$A=-{{1}\over{2\pi\alpha '}}\int \L\; d\sigma_{1}d\sigma_{2},\quad
\L=\sqrt{-\half \L^{\mu\nu}\L^{\mu\nu}},$$
$$\L^{\mu\nu}=\epsilon_{\alpha\beta}\;\partial_{\alpha}x^{\mu}
\partial_{\beta}x^{\nu}\quad \alpha , \beta=1,2$$
(further we fix a system of units $2\pi\alpha '=1$).

The condition for an extremum of action has a form of continuity equation
\begin{equation}
\partial_{\alpha}p_{\alpha}^{\mu}=0,\quad
p_{\alpha}^{\mu}={{\delta A}\over{\delta\;\partial_{\alpha}x^{\mu}}}
={{\L^{\mu\nu}}\over{\L}}\;\epsilon_{\alpha\beta}\;\partial_{\beta}x^{\nu},
\label{cons}
\end{equation}
$p_{\alpha}^{\mu}$ is momentum density, flowing through a section
$\sigma_{\alpha}=Const$.

The integral form of this condition
\eqn{intcons}
{\oint\limits_{C}p_{\alpha}^{\mu}\;\epsilon_{\alpha\beta}\; d\sigma_{\beta}=0}
has a sense of momentum conservation law: total momentum flowing through a
closed contour $C$ on the world sheet, vanishes.

\begin{quotation}
{\small In the Euclidean space analogous
equations give a shape of the soap film surface.
In the theory of soap films the area is proportional to potential energy,
derivatives \rf{cons} are equal to forces of the surface tension. Equation
\rf{intcons} describes an equilibrium of a part of the surface, restricted by
the contour $C$: the total force of tension, acting on this part, vanishes.}
\end{quotation}

One can easily check by substitution, that surface \rf{sheet} obeys condition
\rf{cons}:
$$\partial_{\alpha}x^{\mu}=\half Q'^{\mu}(\sigma_{\alpha}),\quad
\L^{\mu\nu}=\frac{1}{4}\;(Q'^{\mu}(\sigma_{1})Q'^{\nu}(\sigma_{2})
-Q'^{\mu}(\sigma_{2})Q'^{\nu}(\sigma_{1})),$$
$$\L=\frac{1}{4}\;(Q'(\sigma_{1})Q'(\sigma_{2})),\
p_{1}^{\mu}=-\half Q'^{\mu}(\sigma_{2}),\
p_{2}^{\mu}=-\half Q'^{\mu}(\sigma_{1})\quad\Rightarrow\quad
\partial_{\alpha}p_{\alpha}^{\mu}=0.$$

Surface \rf{sheet} has two edges \bfref{straight}. The first edge coincides
with
the supporting curve
$$x_{\mu}(\sigma ,\sigma)=Q_{\mu}(\sigma).$$
The second edge is obtained from the supporting curve by translation onto a
semi-period:
$$x_{\mu}(\sigma +2\pi ,\sigma)=Q_{\mu}(\sigma)+P_{\mu}.$$
The world
sheet transforms into itself in translation on $2P_{\mu}$. When translated
on $P_{\mu}$ it also transforms into itself, in this its edges interchange.

Besides the local equation \rf{cons}, from
the condition of minimum of action it follows
that the momentum flow through the edge vanishes. This requirement holds:
$$\int\limits_{\mbox{{\small along the edge}}}
p_{\alpha}^{\mu}\;\epsilon_{\alpha\beta}\; d\sigma_{\beta}=
\half\int\limits_{(\sigma ,\sigma)}^{(\sigma ',\sigma ')}
Q'^{\mu}(\sigma_{1})d\sigma_{1}-Q'^{\mu}(\sigma_{2})d\sigma_{2}=0.$$

The edges of the world sheet are isotropic: $Q'^{2}=0$. Besides,
any vector from the
tangent plane to the world sheet on its edge is orthogonal to $Q'_{\mu}$.
This plane is tangent to the light cone in the direction of $Q'_{\mu}$.
We will call these planes the isotropic planes.
\begin{quotation}Proof.\quad \small{
Parametrization of world sheet \rf{sheet} is non-regular on the
edge, because tangent vectors ${{\partial x_{\mu}}\over{\partial\sigma_{1}}}$
and ${{\partial x_{\mu}}\over{\partial\sigma_{2}}}$ are linearly dependent
when $\sigma_{1}=\sigma_{2}=\sigma_{0}$
$${{\partial x_{\mu}}\over{\partial\sigma_{1}}}={{\partial x_{\mu}}
\over{\partial\sigma_{2}}}=\half Q'_{\mu}(\sigma_{0}).$$
Let's expand $x_{\mu}(\sigma_{1},\sigma_{2})$ in the vicinity of point
$(\sigma_{0},\sigma_{0})$:
\begin{eqnarray}
x^{\mu}(\sigma_{0}+\Delta\sigma_{1},\sigma_{0}+\Delta\sigma_{2})&=&
Q_{0}^{\mu}+\half Q_{0}'^{\mu}\cdot(\Delta\sigma_{1}+\Delta\sigma_{2})+\nn\\
&&+\frac{1}{4}\; Q_{0}''^{\mu}\cdot({\Delta\sigma_{1}}^{2}+{\Delta\sigma_{2}}
^{2})+o({\Delta\sigma_{1}}^{2}+{\Delta\sigma_{2}}^{2}).\nn
\end{eqnarray}
Regular parametrization in the vicinity of this point is given by
$$\rho_{1}=\Delta\sigma_{1}+\Delta\sigma_{2},\quad
\rho_{2}={\Delta\sigma_{1}}^{2}+{\Delta\sigma_{2}}^{2}\ .$$
Basis vectors of the tangent plane are $Q_{0}'$ and $Q_{0}''$,
\quad$Q_{0}'Q_{0}''=\half(Q'^{2})'_{0}=0,\ $ then all vectors from the tangent
plane are orthogonal to $Q_{0}'$. Equation of $(d-1)$-dimensional tangent space
to light cone $Q'^{2}=0$ in the point $Q_{0}'$ has a form:
$dQ'^{2}\bigr|_{0}=2Q'_{0}dQ'=0$, i.e. vectors $dQ'$ from tangent space are
orthogonal to $Q_{0}'$. The plane specified by vectors $Q_{0}'$ and $Q_{0}''$
is contained in the tangent space to light cone (if $d=3$, these planes
coincide).

Because $\rho_{2}\geq0$, then up to terms $o(\rho_{2})$ the world sheet
lies in a semi-plane of tangent plane, bounded by the straight line in the
direction of
$Q_{0}'$ and containing vector $Q_{0}''$. Therefore, the supporting curve
is the edge of the world sheet.
}\end{quotation}

The edges of world sheet are world lines of string ends. Slice of the world
sheet by constant time plane gives the string at this time moment. The tangent
to the string $l_{\mu}$ is contained in the tangent plane to the
world sheet, that's
why on the edge it is orthogonal to the tangent to the edge
$$l_{\mu}Q_{\mu}'=0\quad\Rightarrow\quad (l_{0}=0)\quad \vec l\vec Q'=0.$$
Therefore, the ends of the string move with light velocity at right angle to
the direction of string in this point.

Total momentum of the string
$$\int\limits_{C}
p_{\alpha}^{\mu}\;\epsilon_{\alpha\beta}\; d\sigma_{\beta}=
\half\int\limits_{(\sigma ,\sigma)}^{(\sigma +2\pi ,\sigma)}
Q'^{\mu}(\sigma_{1})d\sigma_{1}-Q'^{\mu}(\sigma_{2})d\sigma_{2}=P_{\mu}$$
is equal to semi-period of the supporting curve (we have
calculated an integral along a contour, connecting the edges of the sheet).

The projection of the supporting curve into a subspace, orthogonal to $P_{\mu}$
(the rest frame of string) is the closed curve $\vec Q(\sigma)$. The supporting
curve may
be restored by its projection:
$$Q_{0}'=|\vec Q'|\quad\Rightarrow\quad Q_{0}(\sigma_{1})-Q_{0}(0)=
\int\limits
_{0}^{\sigma_{1}}|\vec Q'(\sigma)|d\sigma=L(\sigma_{1}),$$
$L(\sigma)$ is
the arc length of the curve $\vec Q(\sigma)$ between the points $\vec Q(0)$ and
$\vec Q(\sigma)$. The total length of the curve $\vec Q(\sigma)$ is equal to
double mass of string:
$L(2\pi)=2\sqrt{P^{2}}$. One can parametrize the curve $\vec Q(\sigma)$ by its
length:
$\sigma=\pi L/\sqrt{P^{2}}$, then $Q_{0}'=|\vec Q'|=
\sqrt{P^{2}}/\pi$. We will refer to this parametrization as the rest frame
gauge
(RFG).

The shape of string is determined in RFG in the following way
\cite{Pronkopc}. Let's choose an arbitrary point~A on the
 curve $\vec Q(\sigma)$ and
draw two arcs of equal length along the supporting curve
in the opposite sides from point~A and mark the middle of the
segment, connecting these points. A set of all middles forms a curve.
This curve is the string at time moment, when its end is placed in point~A.

In string theory a parametrization is often used
$$\sigma=\pi{{(nQ)}\over{(nP)}},\qquad n_{\mu}\mbox{ is a constant
isotropic vector,}$$
which is called a light cone gauge (LCG). It is convenient to resolve an
equation $Q'^{2}=0$:
$$n=(1,1,..,0),\quad Q_{\pm}=Q_{0}\pm Q_{1},\quad \vec Q_{\perp}=(Q_{2},...,
Q_{d-1}),$$
$$Q'^{2}=Q_{+}'Q_{-}'-\vec Q_{\perp}'^{2}=0\quad\Rightarrow\quad
Q_{-}={{\sigma}\over{\pi}}P_{-},\quad Q_{+}={{\pi}\over{P_{-}}}\int
\vec Q_{\perp}'^{2}(\sigma)d\sigma .$$
If transverse components $\vec Q_{\perp}(\sigma)$ are given in the form of
Fourier expansion, then the last integration can be developed explicitly.

Nevertheless, it occurs that the light cone gauge ``does
 not work'' in space $d=2+1$.
We will return to this question below.

The world sheet (\ref{sheet})
can have singularities. Point of the surface can be
singular, if tangent vectors ${{\partial x_{\mu}}\over{\partial\sigma_{1}}}$
and ${{\partial x_{\mu}}\over{\partial\sigma_{2}}}$ are linearly dependent in
it:
$${{\partial x_{\mu}}\over{\partial\sigma_{1}}}=\half Q_{\mu}'(\sigma_{1})
\quad\|\quad
{{\partial x_{\mu}}\over{\partial\sigma_{2}}}=\half Q_{\mu}'(\sigma_{2})\ .$$
Therefore, for the appearance of singularities on the world sheet
it is necessary, that the supporting curve has two
points with parallel tangents \bfref{onep}.

Tangents are parallel in the points of the supporting curve, separated by the
period. We have found that these points correspond to the edge of the
world sheet.
The edge is the singular line on the world sheet. A surface, obtained by
mapping
(\ref{sheet}) of the whole plane of parameters $(\sigma_{1},\sigma_{2})$ into
Minkowski space, has many coincident sheets -- it repeatedly covers itself,
each point of it is covered with infinite multiplicity. One can easily get
convinced in this, checking that transformations
$$\sigma_{1}\leftrightarrow\sigma_{2};\qquad
\sigma_{1}\to\sigma_{1}+2\pi n,\quad\sigma_{2}\to\sigma_{2}-2\pi n$$
do not change the
point on the surface. In fact, the edge of the surface is a fold,
on which different sheets of this surface are linked. Physically only one
sheet represents the string,
it can be extracted by the restriction of parameters
$(\sigma_{1},\sigma_{2})$ in a band
$\sigma_{1}\leq\sigma_{2}\leq\sigma_{1}+2\pi$.

\begin{quotation}
{\small A surface, consisting of $k$ of such sheets, also represents a possible
evolution of string -- it is a degenerate case of $k$-folded string
\cite{zone}. Degeneration is removed, if one introduces a small deformation on
each $k$-th period of the supporting curve. In this $2P_{\mu}$-periodical curve
becomes $2kP_{\mu}$-periodical, and pleats of $k$ coincident sheets become
disjointed.  }
\end{quotation}

We will look for points with parallel tangents inside one period of the
supporting
curve. In RFG this parallelism condition takes form of equality of unit tangent
vectors to the projection of the supporting curve $\vec Q(\sigma)$:
$$\vec Q'(\sigma_{1})=\vec Q'(\sigma_{2}),\quad
Q_{0}'(\sigma_{1})=Q_{0}'(\sigma_{2})=\sqrt{P^{2}}/\pi .$$
Singularities on the world sheet in spaces $d=3$ and $d=4$ have different
shapes.

\section*{3. Singularities on world sheets in $d=3$ -- kink lines.}

When $d=3$, a space orthogonal to the momentum $P_{\mu}$ is two-dimensional.
Smooth closed curves on a plane are divided into classes with respect to the
winding
number $\nu$ of the unit tangent vector. For convex curves (which do not have
inflection points) a polar angle of the unit tangent vector depends on $\sigma$
monotonously. For each direction of tangent $\nu-1$ points will be found which
have the same tangent direction. Therefore, $\nu-1$ singular points are ever
existing on the string in $d=3$. Singular points on the world sheet
form lines. Tangents to them are isotropic:
$$dx_{\mu}=\half(Q_{\mu}'(\sigma_{1})d\sigma_{1}+
Q_{\mu}'(\sigma_{2})d\sigma_{2})\ \|\ Q_{\mu}'(\sigma_{1}).$$
Let's study a shape of world sheet in the vicinity of a singular point.
\begin{eqnarray}
x(\sigma_{1}+\Delta\sigma_{1},\sigma_{2}+\Delta\sigma_{2})&=&\half(Q_{1}+Q_{2})
+\half Q_{1}'\cdot(\Delta\sigma_{1}+\lambda\Delta\sigma_{2})+\label{singexp}\\
&&+{{1}\over{4}}\;(Q_{1}''\Delta{\sigma_{1}}^{2}+Q_{2}''\Delta{\sigma_{2}}^{2})
+o(\Delta{\sigma_{1}}^{2}+\Delta{\sigma_{2}}^{2}),\nn
\end{eqnarray}
$$Q_{2}'=\lambda Q_{1}'\ .$$
Vectors $Q_{1}',Q_{1}'',Q_{2}''$ are linearly dependent, because $Q_{1}''$ and
$Q_{2}''$ lie in 2-dimensional plane, orthogonal to $Q_{1}'$, and this plane
contains $Q_{1}'$.
$$Q_{2}''=\alpha Q_{1}'+\beta Q_{1}'' ,$$
\begin{eqnarray}
x&=&\half(Q_{1}+Q_{2})
+\half Q_{1}'\cdot(\Delta\sigma_{1}+\lambda\Delta\sigma_{2}+{{\alpha}\over{2}}
\Delta{\sigma_{2}}^{2}))+\nn\\
&&+{{1}\over{4}}\; Q_{1}''\cdot
(\Delta{\sigma_{1}}^{2}+\beta\Delta{\sigma_{2}}^{2})
+o(\Delta{\sigma_{1}}^{2}+\Delta{\sigma_{2}}^{2}).\label{kinkexp}
\end{eqnarray}
Mapping $(\Delta\sigma_{1},\Delta\sigma_{2})\to (X,Y)=
(\Delta\sigma_{1}+\lambda\Delta\sigma_{2},\Delta{\sigma_{1}}^{2}+
\beta\Delta{\sigma_{2}}^{2})$ has a fold on parabola
$Y=X^{2}/(1+\lambda^{2}/\beta)$:
\begin{eqnarray}
&&\det{{\partial(X,Y)}\over{\partial(\Delta\sigma_{1},\Delta\sigma_{2})}}=
2(\beta\Delta\sigma_{2}-\lambda\Delta\sigma_{1})=0\quad\Rightarrow\quad
\Delta\sigma_{2}={{\lambda}\over{\beta}}\Delta\sigma_{1},\nn\\
&&X=(1+{{\lambda^{2}}\over{\beta}})\Delta\sigma_{1}\ ,\qquad
Y=(1+{{\lambda^{2}}\over{\beta}}){\Delta\sigma_{1}}^{2}\ .\nn
\end{eqnarray}
An image of plane $(\Delta\sigma_{1},\Delta\sigma_{2})$ lies over parabola at
$\beta > -\lambda^{2}$ and under parabola at $\beta < -\lambda^{2}$:
$$(1,0)\to(1,1),\quad 1>1/(1+\lambda^{2}/\beta) \mbox{\quad when\quad }
\beta > -\lambda^{2}.$$
In a small vicinity of the point of expansion the world sheet is close to a
semi-plane
with a boundary, parallel to $Q_{1}'$, containing vector $Q_{1}''\mbox{\ sgn}
(\beta+\lambda^{2})$. Therefore, when passing through the singular line,
the world sheet undergoes a fold  \bfref{iso}.
Parts of the world sheet, separated by the singular line,
have common tangent plane near the singular line. This plane is isotropic, as
the tangent plane on the edge.

Slice of the world sheet by constant time plane has a cusp. Such objects
on the string are called kinks. The direction of the tangent to string
in the cusp
(kink direction) is orthogonal to the
kink's world line, because the tangent plane
to the world sheet on the kink line is isotropic. The kink
moves with light velocity
at the right angle to the string direction in this point.

We give a physical explanation
for such behaviour of ends and kinks. We will show
that a mass of vanishingly small part of string adjacent to its end tends
to zero faster than a force of tension acting on this part.

A mass of the small part of string is proportional to its length:
\begin{eqnarray}
dx&=&\half(Q_{1}'d\sigma_{1}+Q_{2}'d\sigma_{2})\quad (dx^{0}=0),\nn\\
dp&=&\half(Q_{1}'d\sigma_{1}-Q_{2}'d\sigma_{2}),\nn\\
(dp)^{2}&=&-\half(Q_{1}'Q_{2}')d\sigma_{1}d\sigma_{2}=-(dx)^{2}={dl}^{2}.\nn
\end{eqnarray}
Let's consider the part of length $\Delta l$, adjacent to the end of string
$(\sigma_{0},\sigma_{0})$:
\begin{eqnarray}
&&\Delta l= \int\limits_{\sigma_{0}}^{\sigma_{0}+\Delta\sigma_{1}}
\sqrt{\half(Q_{1}'Q_{2}'){{Q_{1}^{0'}}\over{Q_{2}^{0'}}}}\; d\sigma_{1}=
\half\sqrt{-Q_{0}''^{2}}{\Delta\sigma_{1}}^{2},\nn\\
&&\Delta\sigma_{1}=-\Delta\sigma_{2}=\sqrt{{{2\Delta l}\over{\sqrt{
-Q_{0}''^{2}}}}}\ .\nn
\end{eqnarray}
The following relations are used
\begin{eqnarray}
&&dx^{0}=0\quad\Rightarrow\quad d\sigma_{2}=-{{Q_{1}^{0'}}\over{Q_{2}^{0'}}}
d\sigma_{1}\ ,\nn\\
&&(Q'Q''')+Q''^{2}=(Q'Q'')'=0\ ,\nn\\
&&(Q_{1}'Q_{2}')=\biggl(Q_{0}'+Q_{0}''\Delta\sigma_{1}+\half Q_{0}'''
{\Delta\sigma_{1}}^{2}+o({\Delta\sigma_{1}}^{2}),\
Q_{0}'+Q_{0}''\Delta\sigma_{2}+\nn\\
&&+\half Q_{0}'''{\Delta\sigma_{2}}^{2}+o({\Delta\sigma_{2}}^{2})\biggr)
=-2Q_{0}''^{2}{\Delta\sigma_{1}}^{2}+o({\Delta\sigma_{1}}^{2})\nn
\end{eqnarray}
Let's obtain a force acting on this part:
\begin{eqnarray}
\Delta p&=&\half\int\limits
_{(\sigma_{0}+\Delta\sigma_{1},\sigma_{0}-\Delta\sigma_{1})}
^{(\tilde\sigma_{0}+\Delta\tilde\sigma_{1},\tilde\sigma_{0}-\Delta\tilde
\sigma_{1})}Q_{1}'d\sigma_{1}-Q_{2}'d\sigma_{2}=\nn\\
&&=\half\biggl(Q(\tilde\sigma_{0}+\Delta\tilde\sigma_{1})
-Q(\sigma_{0}+\Delta\sigma_{1})
-Q(\tilde\sigma_{0}-\Delta\tilde\sigma_{1})
+Q(\sigma_{0}-\Delta\sigma_{1})\biggr)=\nn\\
&&=Q'(\tilde\sigma_{0})\Delta\tilde\sigma_{1}-
Q'(\sigma_{0})\Delta\sigma_{1}=\biggl({{Q'}\over{(-Q''^{2})^{1/4}}}\biggr)'
_{\sigma=\sigma_{0}}\ \sqrt{2\Delta l}\ (\tilde\sigma_{0}-\sigma_{0})\ ,\nn\\
\Delta t &=&Q_{0}^{0'}\ (\tilde\sigma_{0}-\sigma_{0})\ ,\nn \\
F&=&{{\Delta p}\over{\Delta t}}\sim\sqrt{\Delta l},\quad
m\sim\Delta l\quad\mbox{then}\quad F/m\sim1/\sqrt{\Delta l}\to\infty .\nn
\end{eqnarray}
Under such conditions a relativistic particle $m$ moves with light velocity.

\begin{quotation}{\small
The precision of expansions used is insufficient in the points with
zero curvature
$Q''_{0}=0$. In work \cite{Artru} an interesting phenomenon was observed.
Let's consider the supporting curve with straight line segment
$Q'_{1}=Q'_{2}\equiv Q'$. For this curve
$$dx^{0}=0\ \Rightarrow\ d\sigma_{1}=-d\sigma_{2},\qquad
d\vec x=0,\quad dp=Q'd\sigma_{1}\ ,$$
when $d\sigma_{1}$ changes in finite interval. Therefore, a finite isotropic
momentum is concentrated in one point at the end of string.
This momentum accumulates when the end of string passes along the
straight segment
and then flows away up to a moment, when the end of string leaves the straight
segment.
}\end{quotation}

Proof for the kink is analogous. An essential difference between the kink line
 and the edge of the sheet is that the momentum flow through the kink line
does not vanish.
For parts of the world sheet, separated by the kink line, momenta do not
conserve
separately. These parts are not minimal surfaces. We will show that these parts
are causally linked in spite of the fact that metric on the world sheet
degenerates on the kink
line, and all tangent directions to world sheet on kink line are space-like,
except single isotropic direction along kink line.

\subsection*{More about kink direction}

Value $\beta/\lambda^{2}$ is the parametric invariant:
\begin{eqnarray}
&&\sigma\to f(\sigma)\quad \Rightarrow\quad
 Q'\to Q'f',\quad Q''\to Q''f'^{2}+Q'f'',\nn\\
&&\lambda={{Q_{2}^{0'}}\over{Q_{1}^{0'}}}\to \lambda\ f_{2}'/f_{1}'\ ,\quad
\qquad\beta={{(Q_{1}''Q_{2}'')}\over{Q_{1}''^{2}}}\to\beta(f_{2}'/f_{1}')^{2}.
\nn
\end{eqnarray}
In RFG \ $Q^{0''}=0,\ \lambda=1,\ \beta=(\vec Q_{1}''\vec Q_{2}'')/
\vec Q_{1}''^{2}$. Since $\vec Q''\perp\vec Q'$ and $\vec Q_{1}'=
\vec Q_{2}'$, then $\vec Q_{1}''\|\vec Q_{2}''$. Therefore $\beta=\pm
{{|\vec Q_{2}''|}\over{|\vec Q_{1}''|}}$ ($+$, if $\vec Q_{1}''
\uparrow\uparrow\vec Q_{2}''$; $-$, if $\vec Q_{1}''
\uparrow\downarrow\vec Q_{2}''$). $|\vec Q''|$ is proportional to
a curvature of the curve
$\vec Q(\sigma)$: $|\vec Q''|=\left|{{d\ph}\over{d\sigma}}
\right|{{\sqrt{P^{2}}}\over{\pi}}$, where $\ph$ is the polar angle of the
tangent vector
$\vec Q'$. If vectors $\vec Q_{1}''$ and $\vec Q_{2}''$ have the
same direction,
then kink is directed along them. If $\vec Q_{1}''$ and $\vec Q_{2}''$ have
the opposite directions, then the
kink is directed along the vector of a greater module.
Note, that for convex curves $\vec Q(\sigma)$ vectors $\vec Q_{1}''$ and
$\vec Q_{2}''$ have the same direction, because orientation of the
pair $\vec Q',
\vec Q''$ conserves on these curves. This orientation is connected with
orientation of the
supporting curve: the direction of $\vec Q''$ is obtained by rotation
of $\vec Q'$ through $\pi /2$ in the direction of rotation of tangent
\bfref{convex}.

For the
convex supporting curves the kinks move inside the
string never reaching the edges.
(The polar angle of tangent vector monotonously changes with $\sigma$, its rate
${{d\ph}\over{d\sigma}}$ is bounded from above and from below, if the curvature
of the supporting curve never becomes zero or infinity. Therefore, points with
parallel tangents cannot be close as one wishes on the supporting curve,
a ``distance''
$\Delta\sigma$ between such points is separated from zero by positive number.)

The world sheet for the supporting curve \fref{convex} is viewed in
\fref{WStw}.
\begin{quotation}{\small
Density of lines on the left bottom part of this figure is proportional to time
averaged density of the string mass distribution. Points, where density sharply
changes, are placed on the kink line and roundings of strings.
}\end{quotation}
\subsection*{Non-convex supporting curves}

For non-convex supporting curves the direction of rotation of tangent changes
in inflection points. In this new pairs of points with parallel tangents appear
on the curve, new kinks appear on string \bfref{nonconvex}.

The number
of inflection points on closed smooth curves is even. Let's consider the
case, when the supporting curve has two inflection points.

Fig.\ref{newkink} displays the polar angle of tangent vector versus $\sigma$.
On the segments $a,b,c$ for any point $\sigma_{1}$ a point $\sigma_{2}$ exists,
which has the same tangent. These points form curves on the plane of parameters
$(\sigma_{1},\sigma_{2})$:
$$ab=\{ (\sigma_{1},\sigma_{2}):\quad\ph(\sigma_{1})=\ph(\sigma_{2}),\
\sigma_{1}\in a,\ \sigma_{2}\in b\}\quad\mbox{etc.}$$
\begin{quotation}{\small
Let us consider some details of fig.\ref{newkink}. Consider the sections
of a graph $\ph(\sigma)$ by line $\ph=Const$, moving down.
Let $\sigma_{1}<\sigma_{2}$. The equation $\ph(\sigma_{1})=\ph(\sigma_{2})$
begins to have solutions when $\ph <\ph_{A}$. At this moment from the point $A$
the line $ab$ originates. In the vicinity of the point $A$
$$\ph(\sigma_{A}+\Delta\sigma)=\ph_{A}+\half\ph_{A}''
{\Delta\sigma}^{2}+o({\Delta\sigma}^{2}),$$
the solution of equation $\ph(\sigma_{1})=\ph(\sigma_{2})$ in low approximation
is $\sigma_{1}=\sigma_{A}+\Delta\sigma ,\ \sigma_{2}=\sigma_{A}-\Delta\sigma$.
Therefore near point $A$ the line $ab$ is orthogonal to the line $\sigma_{1}=
\sigma_{2}$. In further descending of $\ph$ the value $\sigma_{1}$ decreases
and $\sigma_{2}$ increases until $\ph$ becomes equal to
$\ph_{B}$. At this moment
a rate of $\sigma_{2}$ increases without bound, the rate of
$\sigma_{1}$ remains bounded (the rate of $\ph$ is constant), therefore near
$\ph=\ph_{B}$ the tangent to $ab$ is vertical. In this
$\sigma_{2}=\sigma_{B}$. Now let's lift $\ph$. The value $\sigma_{2}$
passes from interval $b$ to interval $c$. On the line $ac$ values
$\sigma_{1}$ and $\sigma_{2}$ increase until $\ph$ reaches $\ph_{A}$. At this
moment $\sigma_{1}$ has infinite rate, the tangent to $ac$ is horizontal.
In this $\sigma_{1}=\sigma_{A}$. Next we again decrease $\ph$.
The value
$\sigma_{1}$ passes to interval $b$. The line $bc$ terminates in the point $B$.

Replacement $\sigma_{1}\leftrightarrow\sigma_{2}$ gives the whole picture.
}\end{quotation}
In RFG $\sigma\sim Q^{0}$, points of world sheet, satisfying $\sigma_{1}+
\sigma_{2}=Const$, are at the same time in the
rest frame. Value $\beta$ represents
a slope of the tangent to the kink line on the plane $(\sigma_{1},\sigma_{2})$:
$$\vec Q'(\sigma_{1})=\vec Q'(\sigma_{2})\quad \Rightarrow\quad
 \vec Q''(\sigma_{1})
{{d\sigma_{1}}\over{d\sigma_{2}}}=\vec Q''(\sigma_{2}),\ \beta=
{{d\sigma_{1}}\over{d\sigma_{2}}}\ .$$
For convex supporting curves $\beta >0$, then line $\sigma_{1}+\sigma_{2}
=Const$ intersects the kink line in a single point. For non-convex supporting
curves $\beta+1$ can change a sign. For the kink evolution some variants are
possible \bfref{variants}. Kinks appear and disappear lonely on the ends of
string or by pairs inside it. Kinks in pair are oriented to the
opposite sides at
the moment of creating. (The direction of the kink is determined by vector
$\vec Q''_{1}\;\mbox{sgn}(\beta+1)$. Function $\vec Q''_{1}$ is continuous and
does not vanish in the vicinity of the
point C. While passing through the point C the value
$\beta+1$ reverses its sign.) The process of pair creating looks
similarly in any
other reference frame, because the Lorentz transformation does not change time
ordering along the isotropic kink line. The point C of the
world sheet, where a pair
appears, has a complicated structure. In this point $\beta=-1\ (\lambda=1)$,
the precision of expansion (\ref{kinkexp}) is insufficient for the
surface investigation
near this point. Points of single appearance of kinks on the edges also
have a complicated structure. In these points $\beta=-1$ as well.
\begin{quotation}{\small
Non-singular excitations are also evident on figures,
which propagate on the world sheet near non-convex parts
of the supporting curve. These excitations are only the mappings (\ref{sheet})
of non-convex interval $AB$ onto string. However, another interpretation is
possible. One can say that the excitation reflects from the end of string,
during the reflection the non-convex interval on the trajectory of string
end arises. In this the excitation becomes singular -- the kink arises on
string. The kink is absorbed on the end of string and inflects its
trajectory. Besides, the excitations on string can decay into kink pairs.

Our approach is specific because we recover the shape of
the world sheet from the given edge.
In the usual statement the initial data for
Cauchy problem are specified: initial coordinates and momenta of all parts of
string. The supporting curve is defined by this data:
$$Q_{\mu}(\sigma)=x_{\mu}(\sigma)+\int\limits_{0}^{\sigma}d\sigma '
p_{\mu}(\sigma ').$$
The both approaches are equivalent, but they differ in their
interpretation.
}\end{quotation}

Condition of parallelism of tangents conserves in shifts
$$\sigma_{1}\to\sigma_{1}+2\pi n_{1},\ \sigma_{2}\to\sigma_{2}+2\pi n_{2}\ .$$
Images of the kink line in these shifts are shown on \fref{allkinks}. Kinks
appear and disappear on one edge of string, then after a lapse of time, equal
to a semi-period $P_{0}$, these processes repeat themselves
on the another end. These
processes may be overlapped in time -- the new kink can appear before the
disappearance of the
preceding one, a few kinks created by one and the same pair
of inflections can be simultaneously placed on string. But the kink lines
cannot have stable intersections.

Suppose, the kink lines intersect each other in a point $(\sigma_{1}^{*},
\sigma_{2}^{*})$. This implies that the equation $\ph(\sigma_{1})=
\ph(\sigma_{2})$ has two solutions in any vicinity of the point
$(\sigma_{1}^{*},\sigma_{2}^{*})$. Let us expand $\ph(\sigma)$:
$$\ph(\sigma_{1}^{*})+\ph '(\sigma_{1}^{*})(\sigma_{1}-\sigma_{1}^{*})
+\half\ph ''(\sigma_{1}^{*})(\sigma_{1}-\sigma_{1}^{*})^{2}+...=$$
$$=\ph(\sigma_{2}^{*})+\ph '(\sigma_{2}^{*})(\sigma_{2}-\sigma_{2}^{*})
+\half\ph ''(\sigma_{2}^{*})(\sigma_{2}-\sigma_{2}^{*})^{2}+...$$
$$\ph(\sigma_{1}^{*})=\ph(\sigma_{2}^{*})$$
If any of the first derivatives here does not vanish, then the equation can be
unambiguously solved for $\sigma_{1}$ or $\sigma_{2}$. Hence,
the case $\ph '(\sigma_{1}^{*})=\ph '(\sigma_{2}^{*})=0$
is of our interest:
$$\sigma_{2}-\sigma_{2}^{*}=\pm\ \sqrt{{{\ph ''(\sigma_{1}^{*})}
\over{\ph ''(\sigma_{2}^{*})}}}(\sigma_{1}-\sigma_{1}^{*}).$$
The function $\ph(\sigma)$ must have the extrema of the same sign in the
points $\sigma_{1}^{*},\ \sigma_{2}^{*}$. The inflections of the curve $\vec Q
(\sigma)$ are placed in these points. Between the points $\sigma_{1}^{*}$ and
$\sigma_{2}^{*}$ an extremum of the
opposite sign is placed, so the total number
of inflection points must be greater than two. The tangents in the inflection
points $\sigma_{1}^{*}$ and $\sigma_{2}^{*}$ must be parallel to each other,
the kink lines in the point $(\sigma_{1}^{*},\sigma_{2}^{*})$ are also aligned
in this direction. Therefore, a tangency of kink lines occurs in the case
considered, see the central position in \fref{intersection}. This singularity
is not stable. In small deformations of the supporting curve the tangents
in the inflection points become non-parallel, then kink lines disconnect
and have no common points. Pay attention to the fact that
on the right position of
\fref{intersection} the point C appears, where the value $\beta+1$ changes its
sign. In this point the kinks annihilate, later a new kink pair appears in the
point C'.

The kink line on the world sheet can be formed in a closed contour. This takes
place, when the equation $\ph(\sigma_{1})=\ph(\sigma_{2})$ has its solutions
in a vicinity of extrema with the opposite signs \bfref{activity}. If one lifts
the minimum and lowers the maximum, then the contour contracts and disappears
in a moment when  the tangents in the inflection points become parallel
(when the minimum and the maximum are equal).

When $\nu\neq1$,\quad
 $\nu -1$ kinks are constantly present on string. Inflection
points do not significantly affect the
motion of these kinks. An exception is the case,
when in the interval between inflection points A and B the
point $\sigma_{1}$ exists
with the greater negative curvature ${{d\ph}\over{d\sigma}}$, then in
the correspondent point $\sigma_{2}$:
$$\ph(\sigma_{1})=\ph(\sigma_{2}),\qquad -{{d\ph}\over{d\sigma}}(\sigma_{1})>
{{d\ph}\over{d\sigma}}(\sigma_{2})>0.$$
In this case the points with a slope $1/\beta=
{{\ph '(\sigma_{1})}\over{\ph '(\sigma_{2})}}<-1$ are present on the kink line,
 time $x^{0}\sim\sigma_{1}+
\sigma_{2}$ changes non-monotonously along the kink line \bfref{oldkink}.

\subsection*{Notion about light cone gauge in $d=3$}

The following fact shows that LCG works poorly in $d=3$.
Smooth closed curves on a
plane are subdivided into classes with respect to the
winding number $\nu$, which
cannot be continuously deformed to  each other. On the other hand, a shape
of the world sheet in LCG is uniquely determined by function $Q_{2}(\sigma)$
(in space $d=3$ vector $\vec Q_{\perp}$ has only one component), these
functions
form a connected set.

Let's consider a point $\sigma^{*}$ on the supporting curve $\vec Q(\sigma)$,
in
which tangent vector $\vec Q'(\sigma)$ is directed along the
first axis \bfref{LCG}.
In RFG for this point the following relation holds:
$${{dQ_{0}}\over{dL}}={{dQ_{1}}\over{dL}}\ ,\qquad \sigma=\sigma^{*}.$$
But in LCG
$$Q_{0}'-Q_{1}'={{dL}\over{d\sigma}}\left({{dQ_{0}}\over{dL}}-
{{dQ_{1}}\over{dL}}\right)={{P_{-}}\over{\pi}}=Const,$$
that's why ${{dL}\over{d\sigma}}\to\infty$, when $\sigma\to\sigma^{*}$.
In RFG $Q_{0}\sim L$, then function $Q_{0}(\sigma)$ is non-smooth in the point
$\sigma=\sigma^{*}$, function $\vec Q(\sigma)$ is also non-smooth:
\begin{eqnarray}
&&\sigma\to\sigma^{*},\quad Q_{1}'=Q_{0}'-{{P_{-}}\over{\pi}}\to\infty,\nn\\
&&Q_{+}'=2Q_{0}'-{{P_{-}}\over{\pi}}={{\pi}\over{P_{-}}}{Q_{2}'}^{2}\to\infty,
\quad Q_{2}'\to\infty .\nn
\end{eqnarray}
Therefore, LCG parametrizes the supporting curve irregularly in the vicinity of
$\sigma^{*}$. This complicates an analysis of curves smoothness in LCG
parametrization.

The supporting curve has no points with tangent directed along $\vec
e_{1}$, only if $\nu=0$. These curves have at least two inflection points.
The motion of kinks for such world sheets is viewed on \fref{zero}.

For the supporting curves with $\nu\neq0$ LCG parametrization is not regular.
Fourier expansion of non-smooth functions necessarily has an infinite number of
harmonics. Therefore, most of string configurations
are infinite-modal in LCG when $d=3$. Apparently, this makes
string quantization in LCG impossible. Note, that the common obstacle for
string
quantization -- anomaly is absent in $d=3$. Quantization in LCG brings the
anomaly in commutation relations of rotation generators in space, transverse to
momentum \cite{axi}. When $d=3$ this space is two-dimensional, it has only one
rotation generator.

\section*{4. Singularities on world sheets in $d=4$ -- pinch points}

When $d=4$, the space, orthogonal to momentum $P_{\mu}$, is three-dimensional.
In RFG vector $\vec Q'(\sigma)$ lies on a sphere. Singularities on the
world sheet
correspond to points of self-intersection of the curve $\vec Q'(\sigma)$. For
supporting curves in general position singularities on world sheet are placed
in isolated points \bfref{sphere}. (Singular points form kink lines in a
special
case, if the curve $\vec Q'(\sigma)$ is multiple.)

The shape of the world sheet in the vicinity of the
singular point is determined by
expression (\ref{singexp}). In this vectors $Q_{1}',Q_{1}''$ and $Q_{2}''$ can
be linearly independent, because in $d=4$ a space, orthogonal to $Q_{1}'$,
containing these vectors, is three-dimensional.
\begin{eqnarray}
x(\sigma_{1}+\Delta\sigma_{1},\sigma_{2}+\Delta\sigma_{2})&=&
\half(Q_{1}+Q_{2})+\nn\\
&&+\half Q_{1}'\xi+{{1}\over{8\lambda^{2}}}(\lambda^{2}Q_{1}''-Q_{2}'')\xi\eta+
\label{pin}\\
&&+{{1}\over{16\lambda^{2}}}(\lambda^{2}Q_{1}''+Q_{2}'')(\xi^{2}+\eta^{2})
+o(\xi^{2}+\eta^{2}),\nn
\end{eqnarray}
$$Q_{2}'=\lambda Q_{1}',\qquad \xi=\Delta\sigma_{1}+\lambda\Delta\sigma_{2},
\quad\eta=\Delta\sigma_{1}-\lambda\Delta\sigma_{2}.$$
The surface $(X,Y,Z)=(\xi,\ \xi\eta,\ \xi^{2}+\eta^{2})$ is displayed on
\fref{Witny}. The projection of world sheet near the
isolated singular point into any
3-dimensional space looks similarly \bfref{pinch}. Such singular points are
called pinch points. In the pinch point the world sheet
has no tangent plane. As the pinch point is approached along different paths,
tangent planes come to different limiting positions. All these positions lie in
3-dimensional isotropic space, determined by vectors $Q_{1}',Q_{1}'',Q_{2}'$.

At the moment of passing through the pinch point string undergoes
an instantaneous cusp.
\begin{quotation}{\small
In RFG $Q_{1}''^{0}=Q_{2}''^{0}=0,\ \lambda=1$, therefore on the slice of the
surface (\ref{pin}) by the plane $x^{0}=Const$ near the pinch point the
equation $\xi=Const$ holds. In general position vectors
$\vec Q_{1}''-\vec Q_{2}''$ and $\vec Q_{1}''+\vec Q_{2}''$ are linearly
independent. In the coordinate frame $(Y,Z)$, composed by these vectors,
string has a shape of parabola
$$(\xi\eta ,\xi^{2}+\eta^{2})=(Y,\xi^{2}+Y^{2}/\xi^{2}).$$
When the pinch point is approached
$\xi\to0$, the parabola degenerates into a ray along
the positive direction of $Z$ axis. At this moment string has a cusp.
}\end{quotation}

Many interesting properties of pinch points are described
in the remarkable book of G.K.Francis \cite{Francis}. We note here one
property: in any projection on figure plane the pinch point lies on
a contour of the surface, visible or hidden by other parts of the surface.

In the pinch point the lines of self-intersection of the surface terminate.
Note, that the lines of self-intersection are present only on the
projection of the
world sheet in 3-dimensional space. The world sheet shown in \fref{pinch} has
no
self-intersections in 4-dimensional space, because any string forming this
surface does not intersect itself. Stable self-intersections of the world
sheet in 4-dimensional space can be located in isolated points,
as it is shown in \fref{pinp}.

A relation between singularities on sheets in $d=3$ and in $d=4$ is
demonstrated
by the following puzzle. Curves with different winding numbers $\nu$ cannot be
transformed to each other by the continuous deformation on a plane. They can be
transformed to each other by ``twisting through the third dimension''
\bfref{puzzle}. What will happen with the world sheet in such deformation?
The answer is given in \fref{answer}.

\begin{quotation}\small
As soon as the supporting curve $\vec Q(\sigma)$ becomes non-flat, the
kink line is
smoothed down. The projection of the world sheet into rest frame intersects
itself along the line $PQ$, beginning in pinch point $Q$
and terminating on the edge of the sheet.
In further deformation the pinch point moves to the edge of the sheet and
disappears there. Then the
supporting curve becomes flat with $\nu=1$, the world sheet
has no singularities.

In the inverse order:\ the
sheet turns up on the edge, the double line and the pinch point
appear. Then the surface is rumpled along kink line.
\end{quotation}
Small deformations of the
supporting curve into the third dimension can create a number
of pinch points. In described deformation the pinch points
disappear in pairs, as is
shown in \fref{ansp} (compare with fig.9,10 in \cite{Francis}).

\subsection*{Notion about light cone gauge in $d=4$}

RFG and LCG parametrizations are connected by stereographic
projection \bfref{stereo}:
$$\vec Q_{\perp}'^{(LCG)}=\left({{d\vec Q}\over{dL}}\right)_{\perp}{{dL}\over{
d\sigma}}=
\left({{d\vec Q}\over{dL}}\right)_{\perp}{{P_{-}/\pi}\over{1-{{dQ_{1}}
\over{dL}}}}\ ,$$
\begin{eqnarray}
&&{{BD}\over{CE}}={{AB}\over{AE}}\ ,\nn\\
&&BD=|\vec Q_{\perp}'^{(LCG)}|,\qquad CE=\left|\left({{d\vec
Q}\over{dL}}\right)
_{\perp}\right|\ ,\nn\\
&&AO=1,\qquad AE=1-{{dQ_{1}}\over{dL}}\ ,
\qquad AB={{\sqrt{P^{2}}}\over{\pi}}\ ,\nn
\end{eqnarray}
$P_{-}=P_{0}=\sqrt{P^{2}}$ for $n_{\mu}=(1,1,0,0)$ in the reference frame,
connected to momentum $P_{\mu}=(\sqrt{P^{2}},0,0,0)$.

When $d=4$, supporting curves also exist which have a tangent
$\vec Q'^{(RFG)}(\sigma)$,
directed along $\vec e_{1}$ in some point.
Alignment of $\vec Q'(\sigma^{*})$ along $\vec e_{1}$ can be removed by small
deformation of the curve. Positions $\vec Q'(\sigma^{*})\uparrow\uparrow
\vec e_{1}$ are not topological obstacles for deformations, their presence does
not lead to the division of curves into non-equivalent classes. Smooth curves
$\vec Q(\sigma)$ in RFG
correspond to continuous curves $\vec Q'(\sigma)$ on a sphere.
Smoothness of curve $\vec Q'(\sigma)$ can be violated (\fref{violate}).
Non-smoothness of the curve $\vec Q'(\sigma)$ does not necessarily mean
the non-smoothness of
correspondent functions in LCG and infinite number of modes in
their Fourier expansions: not more than $\vec Q''=0$ is in this point.

All continuous curves on a sphere can be transformed to each other by
continuous
deformation. These deformations can be developed
on a sphere with the rejected point $\vec e_{1}$. This can require large
deformations. Initial and final positions of the
curve on \fref{closeRFG} are close
in RFG: $|\vec Q'(\sigma)_{i}-\vec Q'(\sigma)_{f}|<\varepsilon$, but in LCG
they are not close (in the same sense), see \fref{closeLCG}.

\section*{5. Characteristic lines and causal structure of the world sheet}

When the supporting curve $Q_{\mu}(\sigma)$ is non-smooth in a
point $\sigma_{0}$,
the world sheet has angular points along a line
\eqn{gen}{x_{\mu}(\sigma_{0},\sigma)=\half(Q_{\mu}(\sigma_{0})+Q_{\mu}(
\sigma)).}
This line is proportional to the supporting curve with coefficient $1/2$
(\fref{induce}). Tangent planes to the world sheet on the fracture line are
determined by vectors
$(Q_{\mu}'(\sigma_{0}-o),Q_{\mu}'(\sigma))$ and $(Q_{\mu}'(\sigma_{0}+o),
Q_{\mu}'(\sigma))$, these planes do not coincide. Inside the
regions of the world sheet,
separated by line (\ref{gen}), condition (\ref{cons}) holds. On line
(\ref{gen}) the conservation law holds: a
momentum flowing from one part of the sheet
is equal to that flowing into another part and is equal to
$$\half\int\limits_{\mbox{{\small along line (\ref{gen})}}}
Q'(\sigma)d\sigma.$$
Therefore, this piecewise smooth surface is minimal.
\begin{quotation}
{\small If higher derivative $Q^{(n)}$ is subjected to discontinuity in the
point
$\sigma_{0}$, then the world sheet has discontinuity in derivatives of the
same order
on the line (\ref{gen}). If a parametric invariant of the curve $Q(\sigma)$ is
available, which depends on derivatives up to $Q^{(n)}$ and undergoes
discontinuity in the
point $\sigma_{0}$ (i.e. not only parametrization of the curve is
non-regular in the point $\sigma_{0}$, but the curve itself is non-regular),
then the world sheet
has a parametric invariant depending on derivatives of the
same order and undergoing
discontinuity along line (\ref{gen}). For example, if the curvature of curve
$\vec Q(\sigma)$ has discontinuity in point $\sigma_{0}$, then invariants of
the
second quadratic form of world sheet have discontinuity on line
(\ref{gen}).}
\end{quotation}

Therefore, irregularities of the supporting curve propagate on the
world sheet along
isotropic lines (\ref{gen}).
Parameters $(\sigma_{1},\sigma_{2})$ form isotropic coordinates on
the surface:
$(\partial_{1}x)^{2}=(\partial_{2}x)^{2}=0$.
One can easily show that equation (\ref{cons}) in these
coordinates takes a form $\partial_{1}\partial_{2}x=0$.
The lines (\ref{gen}) are characteristics of equation.
This is in agreement with a concept of {\it genes}, introduced in work
\cite{Artru}, which implies that the
irregularities of initial data propagate on
the world sheet along characteristics.

Characteristics, edges and kink lines exhaust all isotropic
lines on the world sheet:
$$(Q_{1}'\dot\sigma_{1}+Q_{2}'\dot\sigma_{2})^{2}=0\ \Rightarrow\
(Q_{1}'Q_{2}')\dot\sigma_{1}\dot\sigma_{2}=0,$$
\begin{itemize}
\item $\dot\sigma_{1}\dot\sigma_{2}=0:$\\
$\sigma_{1}=Const\mbox{\ or\ }\sigma_{2}=Const$\quad characteristics\\
(through each regular point of the surface two characteristics pass);
\item $(Q_{1}'Q_{2}')=0:$\\
$\sigma_{2}=\sigma_{1}\mbox{\ or\ }\sigma_{2}=\sigma_{1}+2\pi$\quad the
edges.\\
In other cases\ $\vec Q_{1}'\vec Q_{2}'=|\vec Q_{1}'||\vec Q_{2}'|
\ \Rightarrow\ Q_{1}'\| Q_{2}'$\quad the kink line.
\end{itemize}

Characteristic marks the path of a light signal along the
world sheet. Two characteristics pass through each inner point of the world
sheet. Single characteristic passes through a point on the edge.
Characteristics that pass through kink lines and edges are tangent
to these curves \bfref{char} (they are tangent
in Minkowski space, but not on the parameters plane).

\subsection*{Causal structure}

Characteristics introduce a causal structure on the world sheet. This structure
determines, whether two points on the world sheet communicate by a signal,
propagating along world sheet with a velocity, not greater
than the light velocity.
Characteristics, passing through a given point, divide the world sheet into
a future region, a past region and regions of causal independence \bfref{caus}.
The singularities do not affect causal structure (the kink line is an obstacle
for signals, the velocity of which is strictly less than the velocity of light,
but is transparent for light signals).

Points of self-intersections make the causal structure to be more complex
(the right part of \fref{caus}). The point of self-intersection is marked by
two separate points $S$ and $S'$ on the parameters plane, which are glued
in Minkowski space. If the self-intersection point $S$ is placed in the
future region for point $A$, then the consequences of $A$ are also $S'$,
the future region for $S'$, the points of self-intersections, contained
in $F'$ etc.

\subsection*{Position of singular world sheets among other surfaces}
The world sheet with singularities is not a special case of the world sheet,
but on the
contrary, it is a world sheet ``in general position''. This position is quite
analogous to the position of smooth non-monotonic function in a set of all
smooth functions. A set of singular
world sheets, endowed by the appropriate definition
of world sheets closeness\footnote{For example, uniform closeness of
derivatives $\vec Q'(\sigma)$ in RFG, see the end of Section~4.},
forms a region. For the sheets inside the region
the singularities cannot be eliminated by small deformation. Sheets without
singularities lie on the boundary of the region. This means that in any
vicinity of a sheet without singularities the singular world sheets are
available, i.e. singularities are created by small deformation.

For general surfaces the singular points considered are not stable. The folds
of surfaces in 3D space are eliminated by smoothing. For example, a half-cubic
fold is eliminated by homotopy
$(\xi ,\eta^{2},\eta^{3})\to(\xi ,\eta^{2},\eta^{3}+\epsilon\eta)$.
The pinch points are stable in 3D space, but are not stable in 4D.
The surface (\ref{pin}) lies in 3-dimensional subspace of 4-dimensional space
(up to $o(\xi^{2}+\eta^{2})$).
Homotopy $(\xi ,\xi\eta ,\xi^{2}+\eta^{2},0)\to(\xi ,\xi\eta ,
\xi^{2}+\eta^{2},\epsilon\eta)$ eliminates the singular point on this surface
(for all $\epsilon\neq0$ the tangent vectors to this surface are linearly
independent).

Therefore, kink lines in 3D
and pinch points in 4D are stable on minimal surfaces, but can be eliminated
by deformation, violating the minimal property. These singular points
are specific namely for minimal surfaces. The singularities induced by
non-smoothness of the
supporting curve are not stable (they can be eliminated by
smoothing). The lines of self-intersection in 3D and points of
self-intersections in 4D are stable for all surfaces.

There are no stable singular points on the world sheets in the space with
$d>4$, because in this case the curve $\vec Q'(\sigma)$ lies
on $(d-2)$-dimensional sphere, and its self-intersections can be eliminated
by small deformations.

\section*{6. Outlook}

We have studied the singularities on the world sheets of open string.

When $d=3$, singularities have the form of kinks, propagating on string with
the light velocity. If projection of the supporting curve into the rest frame
is
the convex curve with
winding number $\nu$, then the number of kinks conserves and equals
$\nu-1$. If this curve is non-convex, then new kinks periodically appear and
disappear on string. Kinks appear and disappear lonely on the ends of
string or in pairs inside it.

When $d=4$, the singularities on the world
sheet are placed in isolated pinch points.
The string undergoes an instantaneous cusp at the moment of passing
through the pinch point.

When $d>4$, there are no stable singularities on the world sheets.

If the supporting curve itself has singular points, then the
correspondent singularities
on the world sheet are placed on characteristics, passing through these points.

We will study the singularities on the world
sheets of other topological types and singularities in processes of string
breaking elsewhere.

\subsection*{Acknowledgements}
We are pleased to thank George P. Pron'ko,  Tosiyasu L. Kunii and Hans Hagen
for fruitful discussions.
We are gratefully acknowledge Russian Ministry of Science for the support
of this work.

\vspace{0.5cm}

\hfill {\it Received December 9, 1994.}

\newpage
\section*{Captions to figures}

\fig{straight}{*}
{World sheet of string is constructed as
geometrical place of middles of segments,
connecting all possible pairs of points on the supporting curve.
It has two isotropic edges: the supporting curve (1) and its image
in translation
onto the semi-period (2). A slice of the world sheet by the plane of
constant time gives string in this instant of time. Time axis is directed
from left to right.}

\fig{onep}{}
{A point on a surface is regular, if tangent vectors are linearly independent
in it. Otherwise the point is singular. The numbers on this figure denote:
1 -- regular point, 2,3 -- singular points. The point 2 lies on the edge of
the world sheet.}

\fig{iso}{}
{The shape of the world sheet near singular line.}

\fig{convex}{*}
{Kink on string (in the rest frame). The numbers on the figure denote:
1 -- supporting curve, 2 -- trajectory of the kink, 3 -- the string.
Direction of the kink $\vec Q''$, direction of its velocity $\vec Q'$
and direction of motion of the string ends are related (see text).}

\fig{WStw}{*}
{Left upper part shows the world sheet for the supporting curve, shown
on the previous figure. Left bottom part shows the projection of this
world sheet into the rest frame. Numbers denote: 1,2 -- the edges of the sheet,
3 -- kink line. Note, that near a point $Q$ the edge 1 passes at first
in front of the kink line, then behind it. Near the point $Q$ the world sheet
has self-intersection. Right part displays (from top to bottom):
the kink, the line of self-intersection $PQ$, the self-intersection
of string.}

\fig{nonconvex}{}
{Non-convex supporting curve and its unit tangent vector. A,B -- inflection
points. The tangents in 1,2,3 are collinear, therefore additional kink lines
appear on string.}

\fig{newkink}{}
{Polar angle of tangent vector dependence of $\sigma$
and the kink line on a plane of parameters $(\sigma_{1},\sigma_{2})$.
$a,b,c$ -- the intervals of monotonity of the function $\ph(\sigma)$.
1,2,3 -- the points with parallel tangents, see previous figure.}

\fig{variants}{*}
{The world sheets for non-convex supporting curves. Left upper parts show the
kink line on the plane of parameters $(\sigma_{1},\sigma_{2})$.
\begin{description}
\item\quad\quad
a)\ Kink appears on the end of string at the moment, when the end passes
through the inflection point A. Kink disappears on the same end when it
passes through the inflection point B. Between the points A and B
the world sheet intersects itself along the line $PQ$.
\item\quad\quad
b)\ At some moment in the point C two kinks appear on string. One kink
disappears on the end of string in the point A, another one disappears in the
point B.
\item\quad\quad
c)\ In the point C the pair of kinks appears. One kink disappears on the end of
string in the point A. In the point B new kink appears, it annihilates
the second kink in the point C'.
\end{description}}

\fig{allkinks}{} {Images of the kink line in translations
$\sigma_{1}\to\sigma_{1}+2\pi n_{1},\ \sigma_{2}\to\sigma_{2}+2\pi n_{2}$,
which do not change the parallelism of tangents. The kinks appear and disappear
on one end of string, then after a lapse of time equal to the semi-period
$P_{0}$ these processes are repeated on another end.}

\fig{intersection}{*} {Rearrangement of kink lines. In deformation of the
supporting curve the kink lines approach each other, become tangent, then the
scattering of the kinks transfers to ``the annihilation channel''.}

\fig{activity}{*} {Closed kink line on the world sheet. The kink pair appears
in
the point C and disappears in the point C'. The supporting
curve has 4 inflection points.}

\fig{oldkink}{*} {The world sheet for the non-convex supporting curve
with $\nu=2$.
One kink (a) is constantly present on string. At some moment in the point C
the kink pair appears, one kink of which annihilates kink (a) in the point C'.
Another kink continues to move on string.}

\fig{LCG}{} {In the point $\sigma^{*}$ tangent vector to the supporting curve
is
directed along axis~1. Function $Q_{0}(\sigma)$ has an infinite derivative in
this point.}

\fig{zero}{*} {
\begin{description}
\item Upper figure.\quad
Supporting curve with $\nu=0$. Its unit tangent vector has two cusps A and B,
correspondent to the inflection points on the supporting curve. Polar angle
$\ph(\sigma)$.
\item Bottom figure.\quad The world sheet and a shape of string. Solutions of
equation $\ph(\sigma_{1})=\ph(\sigma_{2})$ on the plane
$(\sigma_{1},\sigma_{2})$. Their images in shifts
$\sigma_{1}\to\sigma_{1}+2\pi n_{1},\ \sigma_{2}\to\sigma_{2}+2\pi n_{2}$.
The kink appears at one end of string and disappears on another end.
\end{description}}

\fig{sphere}{} {The world sheet in $d=4$ has singularities,
when a curve $\vec Q'(\sigma)$ has points of self-intersection.}

\fig{Witny}{*}
{The world sheet in the vicinity of isolated singular point.
Axes form a basis in some 3-dimensional isotropic space
(tangent to light cone). }

\fig{pinch}{*} {Projection of the world sheet into the
rest frame near pinch point
$Q$. This surface intersects itself along the curve $PQ$, terminating in the
pinch point. String has a cusp in the pinch point.}

\fig{pinp}{}
{Self-intersection of string.}

\fig{puzzle}{} {Smooth closed curves with different winding numbers
are not topologically equivalent on the plane, but are equivalent in 3D space.}

\fig{answer}{*}{The deformation of the world sheet,
correspondent to the deformation of
the supporting curve on previous figure. Near each world sheet the curve
$\vec Q'(\sigma)$ is also displayed.}

\fig{ansp}{}
{Annihilation of the pinch points.}

\fig{stereo}{}
{RFG and LCG parametrizations are related by stereographic projection.}

\fig{violate}{} {The curves $\vec Q'(\sigma)$ are continuous. Smoothness of
$\vec Q'(\sigma)$ can be violated in the deformation of smooth curves
$\vec Q(\sigma)$. This happens when the pinch point disappears on the edge, see
\fref{answer}.}

\fig{closeRFG}{} {Two close curves i and f can be transformed one to another by
deformation on a sphere with rejected point.}

\fig{closeLCG}{} {The deformation, displayed on previous figure,
on the plane $\vec Q'_{\perp}$ in LCG.}

\fig{induce}{*} {Fracture of the world sheet, caused by non-smoothness of
the supporting curve. Line of fracture is obtained from the supporting curve by
contraction in 2 times to point $\sigma_{0}$.}

\fig{char}{*} {Characteristics are tangent to the edges and the kink line. Grid
on the world sheet is composed of the characteristics. All characteristics
are proportional to the supporting curve with the coefficient 1/2 and therefore
are congruent. The
edges 1,2 and the kink line $k$ are marked. The kink line divides the world
sheet into two parts. The marked characteristic passes from one part to another
through the kink line and remains smooth. Time required for propagation of
the characteristics from one edge to another is constant and equals
$P_{0}$.}

\fig{caus}{}
{Causal structure on the world sheet. $F$ -- future region, $P$ -- past region,
$I$ -- causally independent regions.}

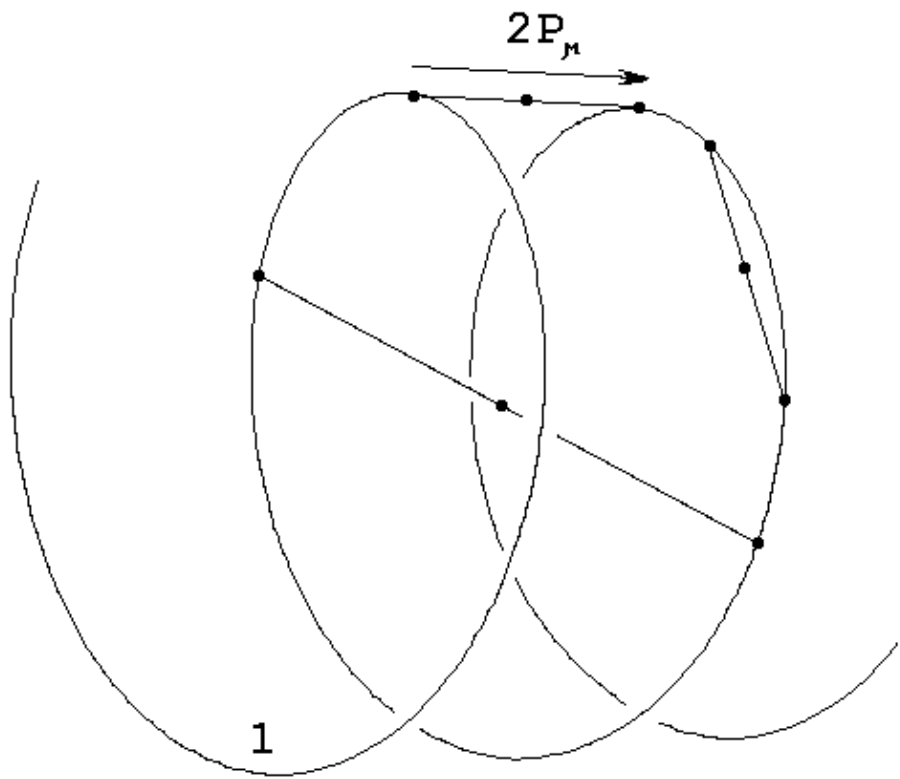
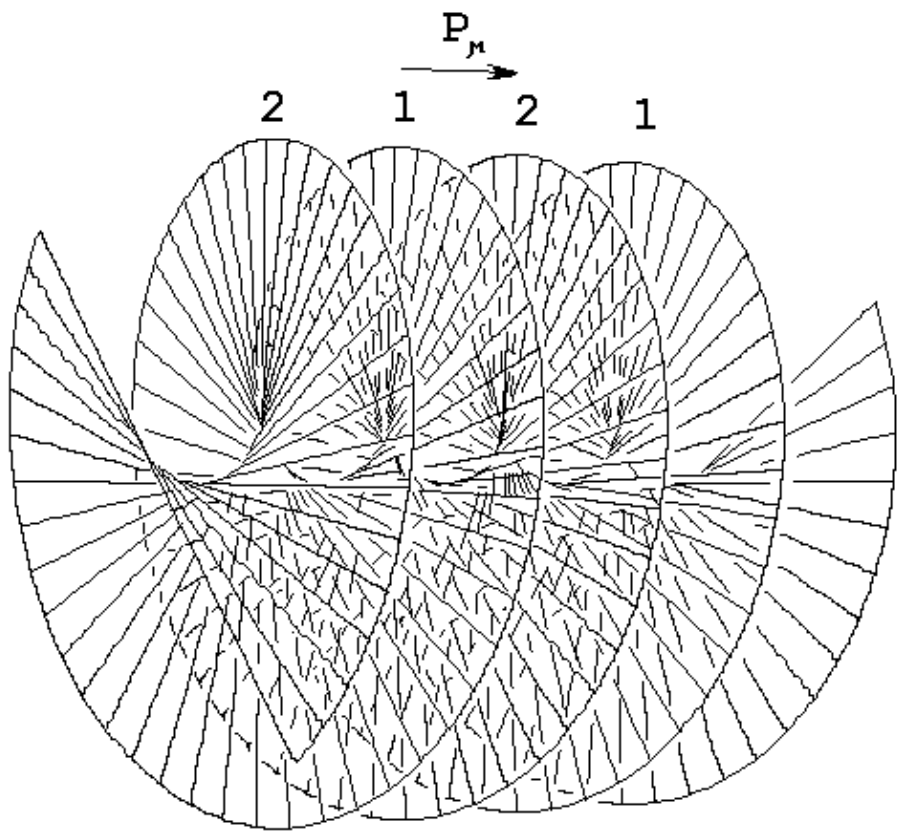
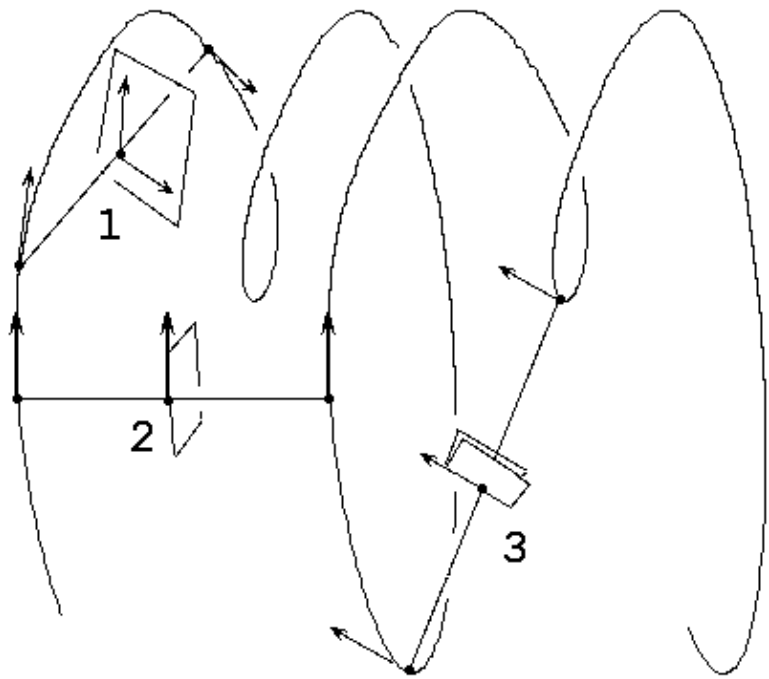
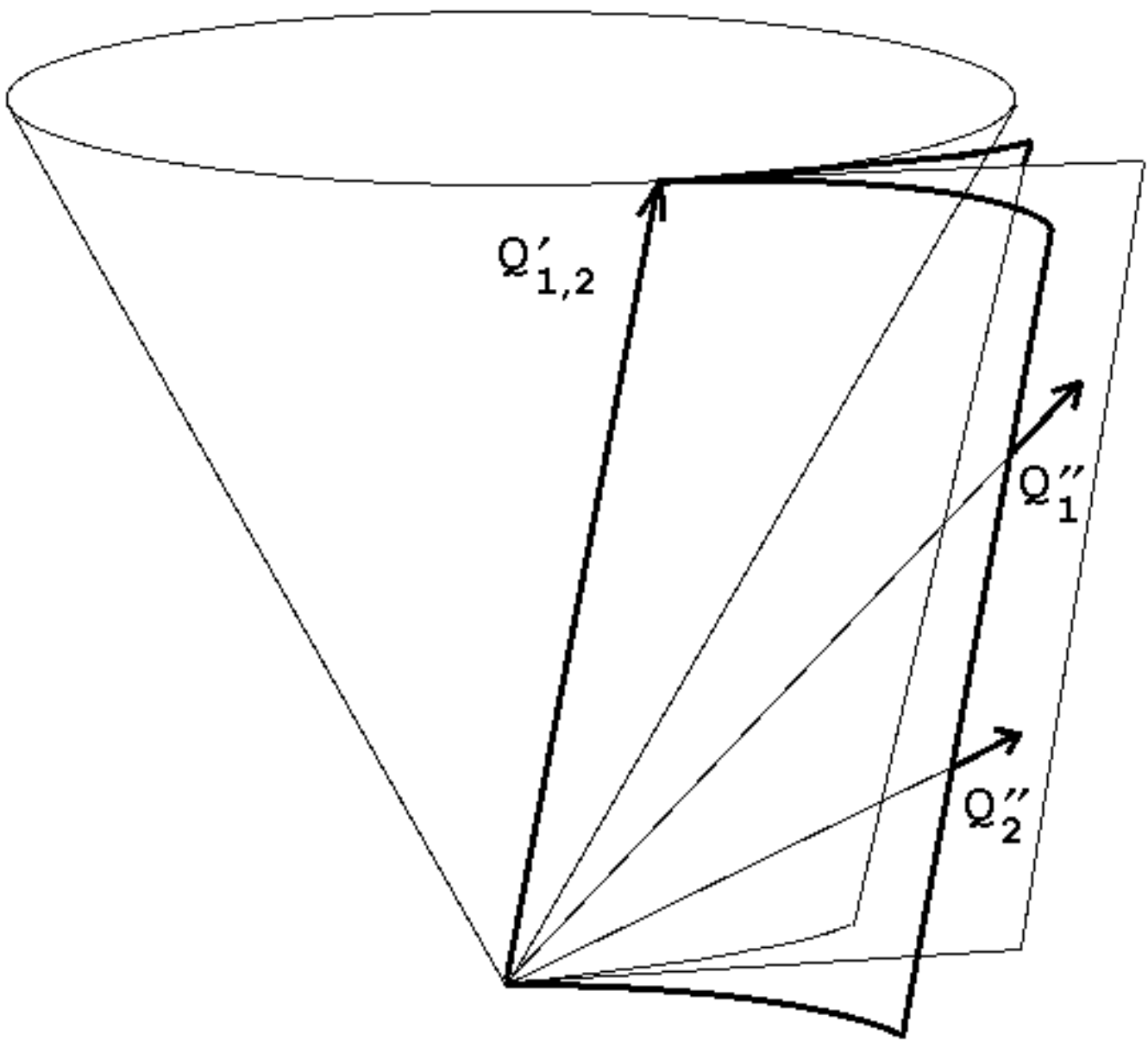
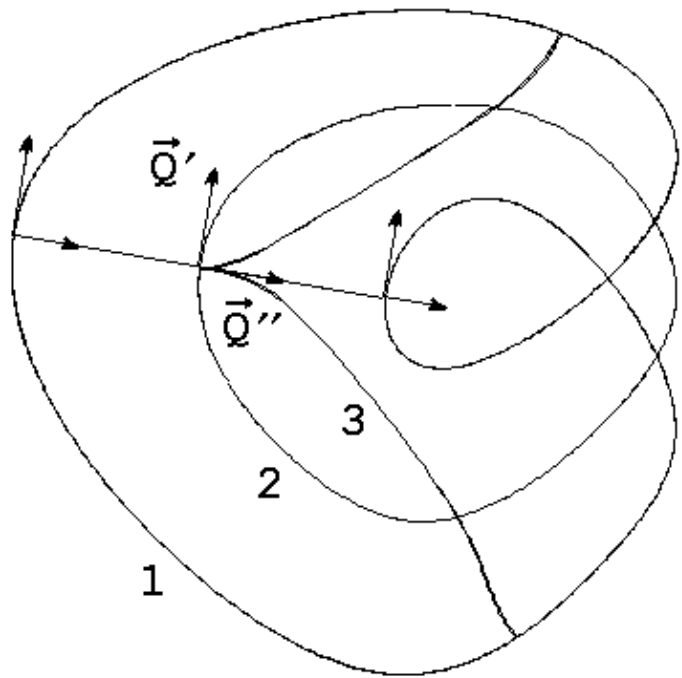
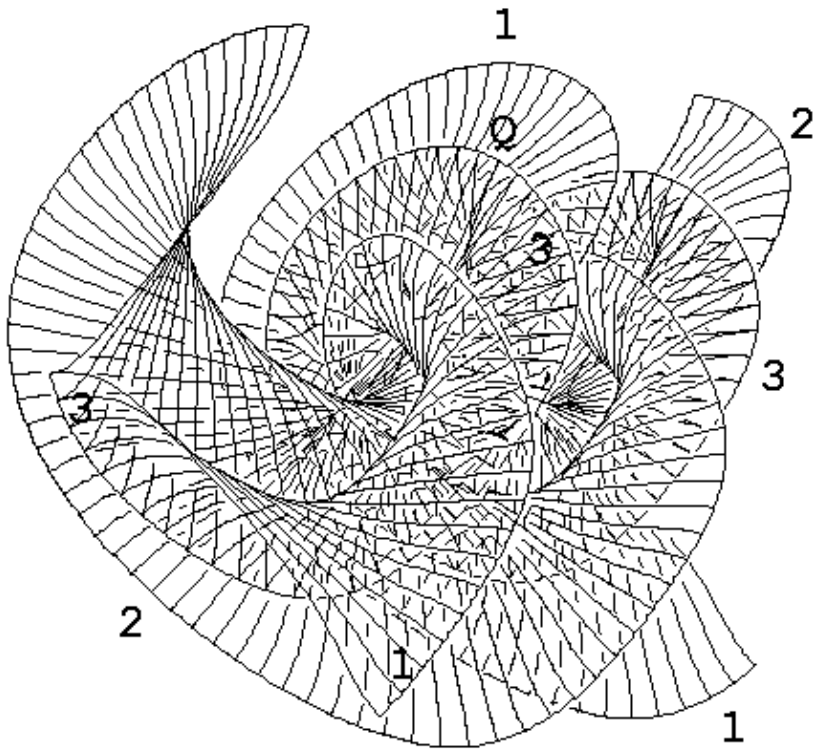
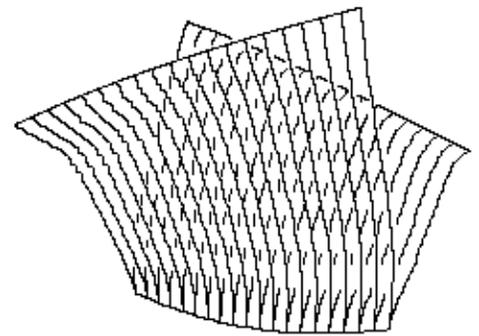
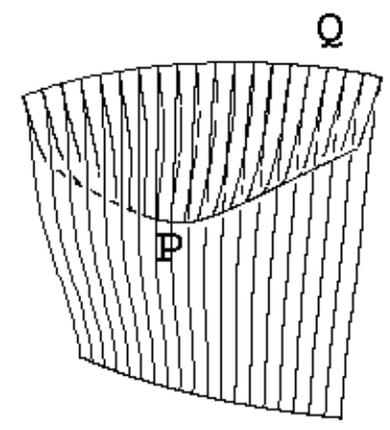
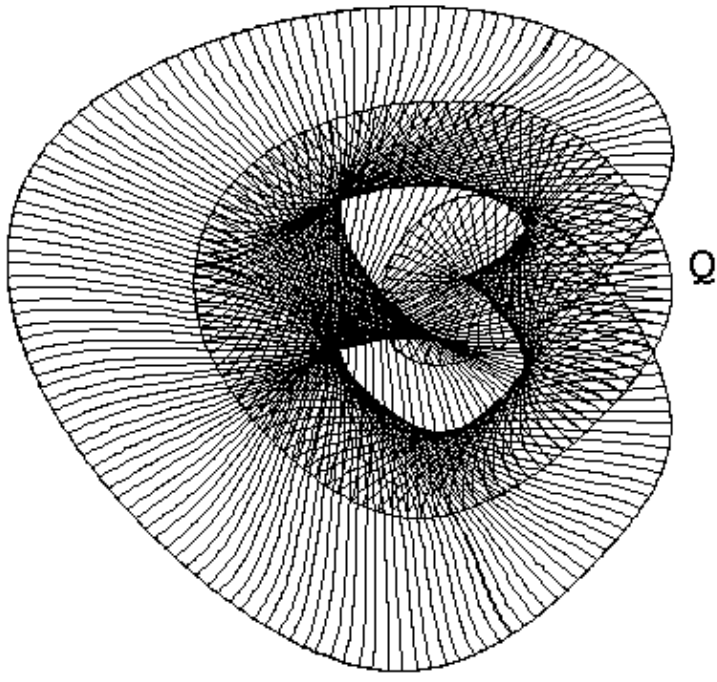
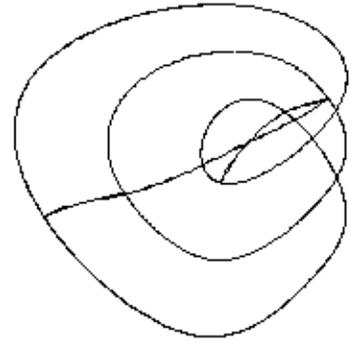
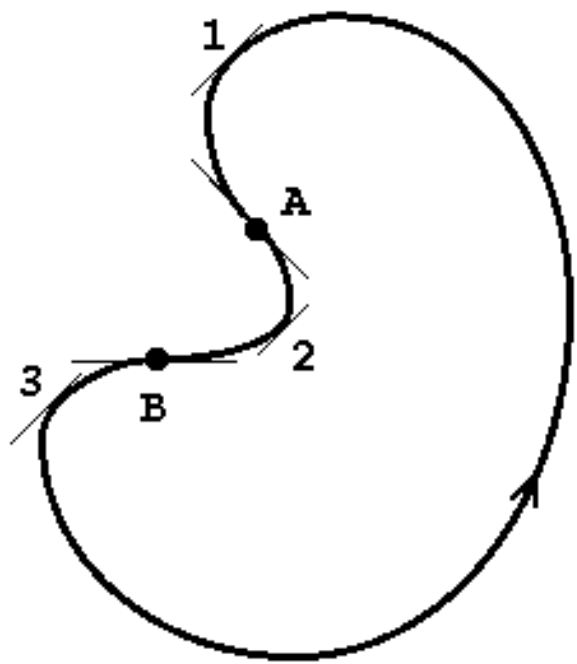
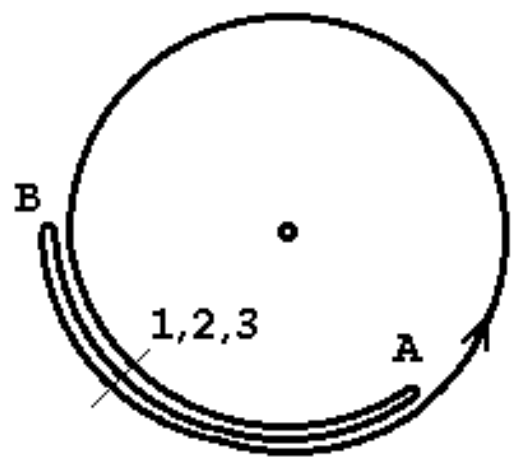
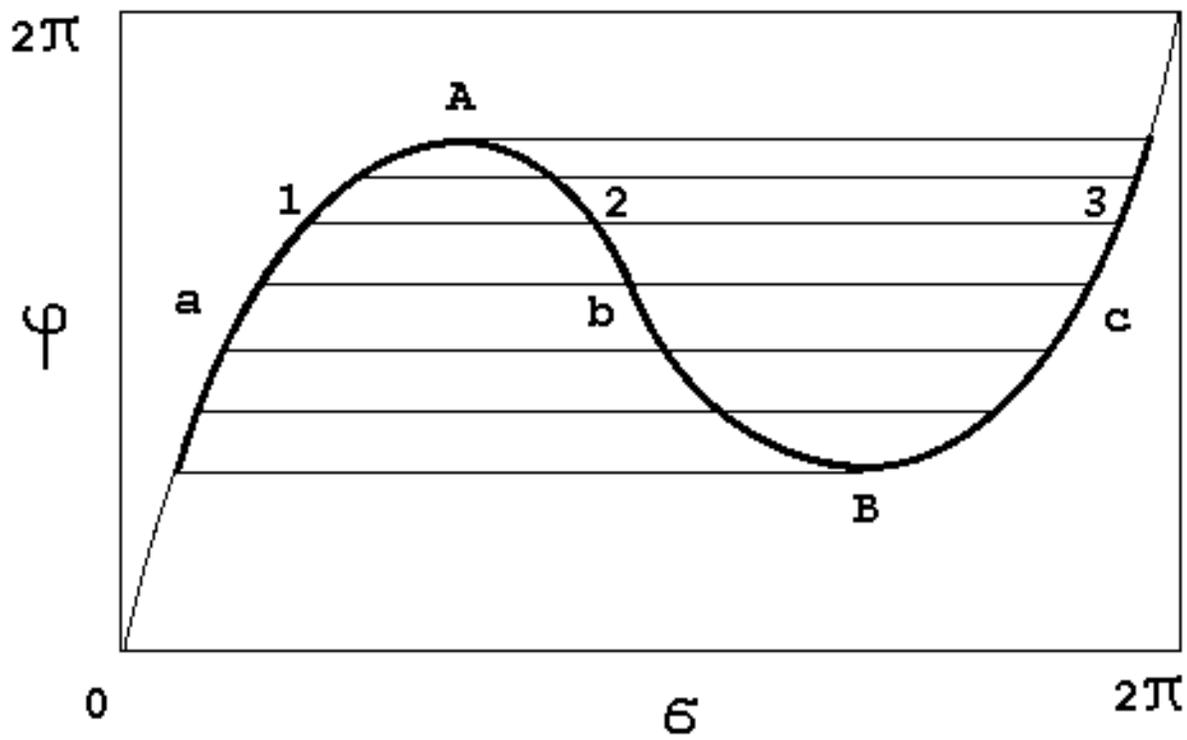
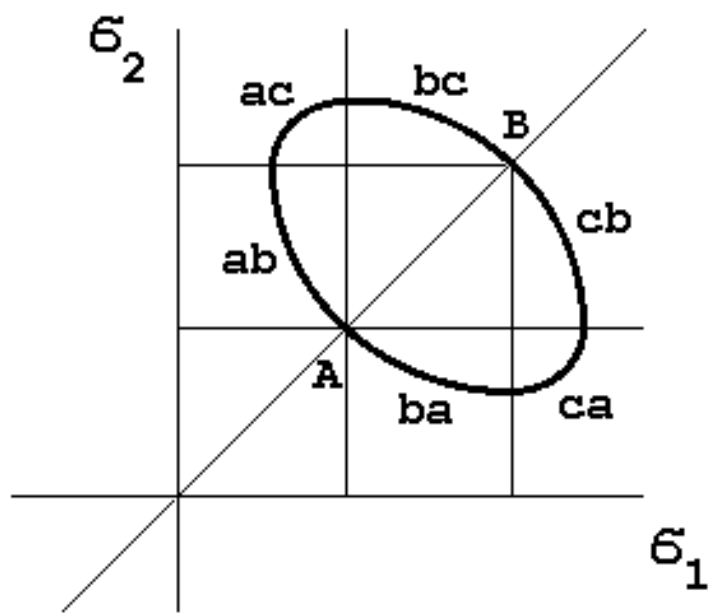
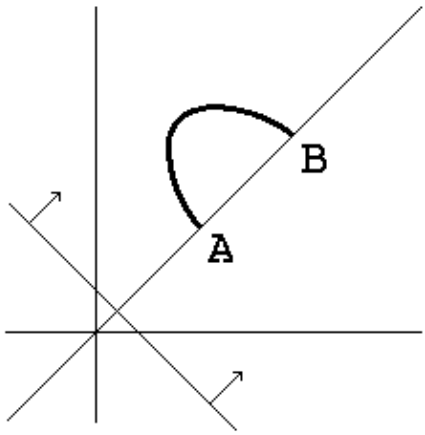
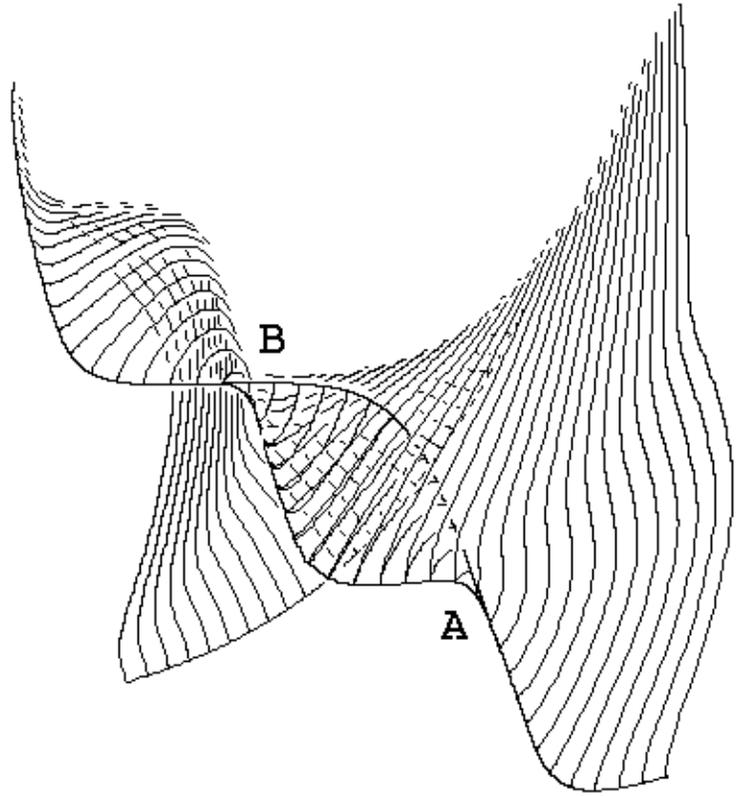
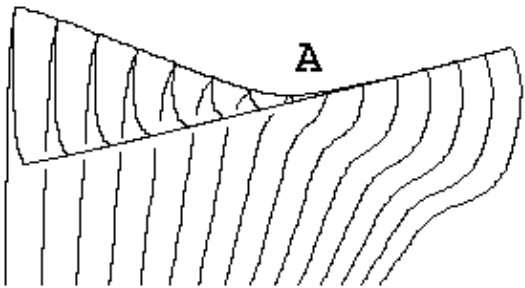
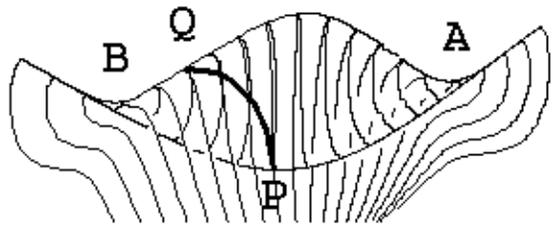
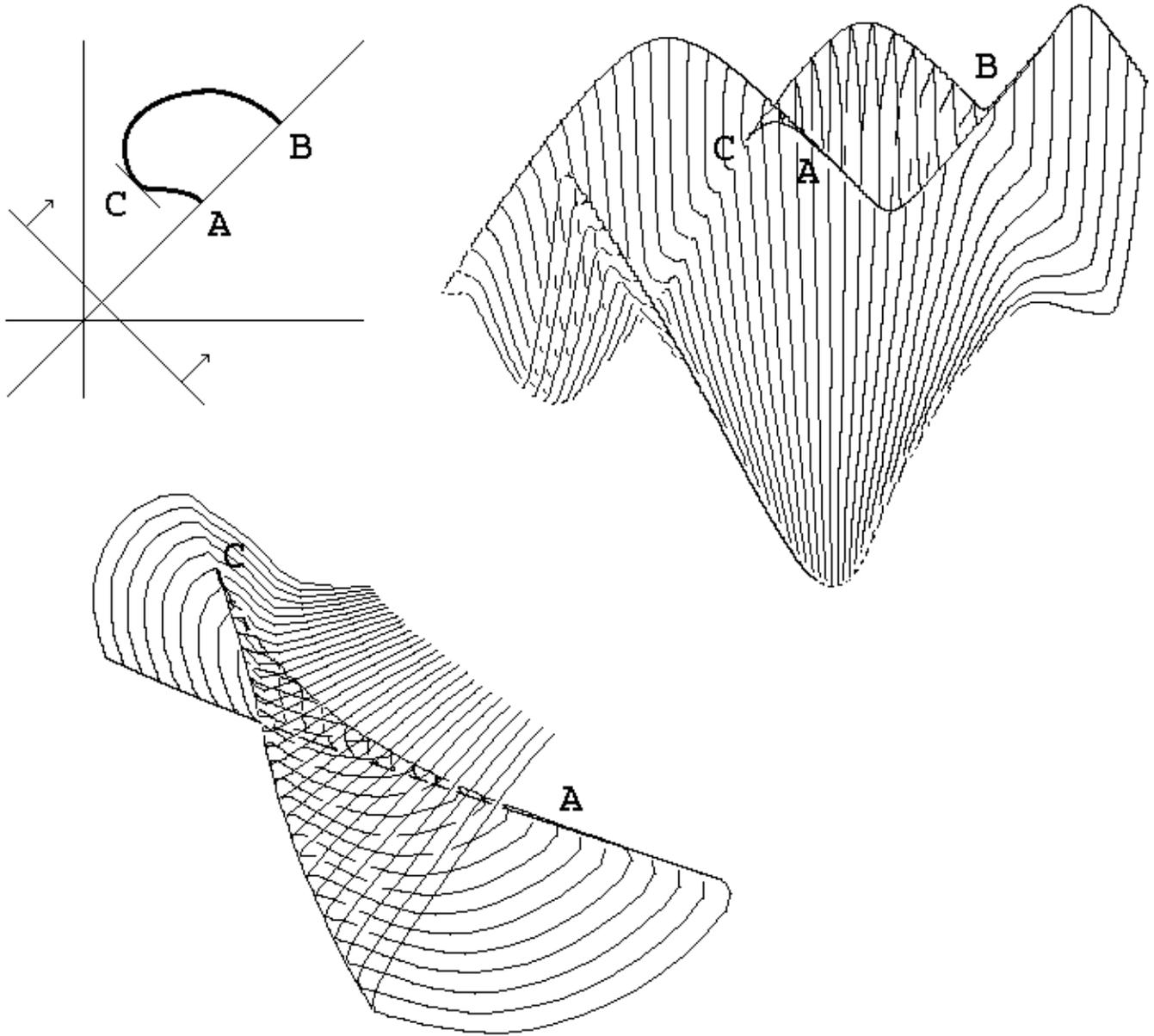
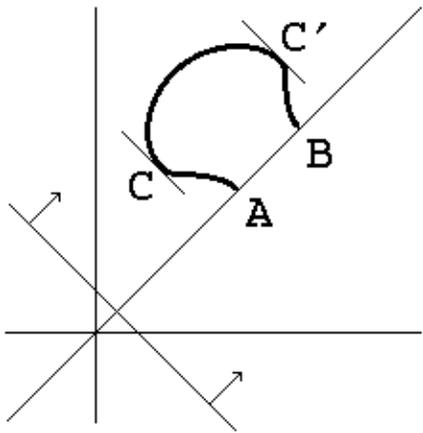
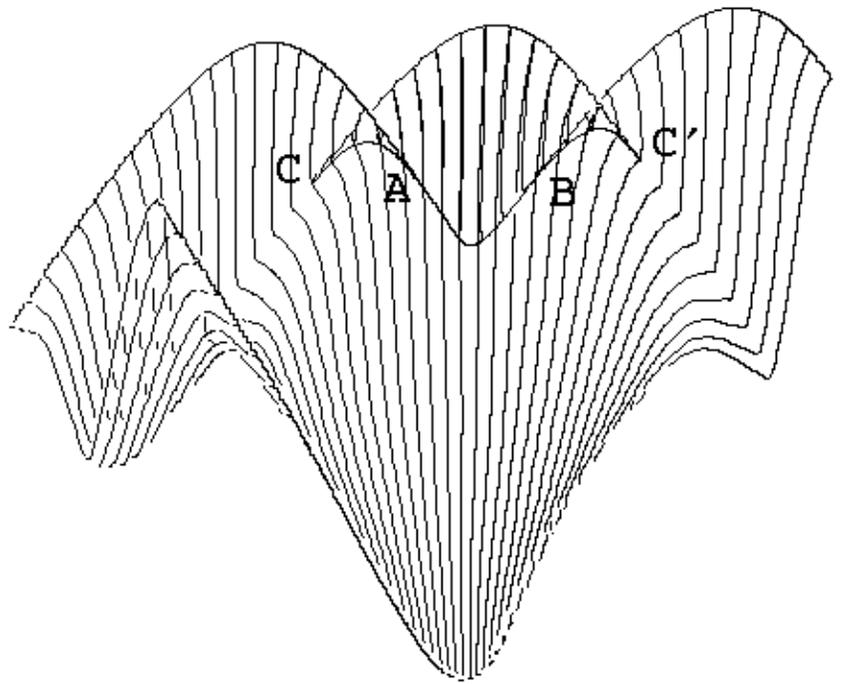
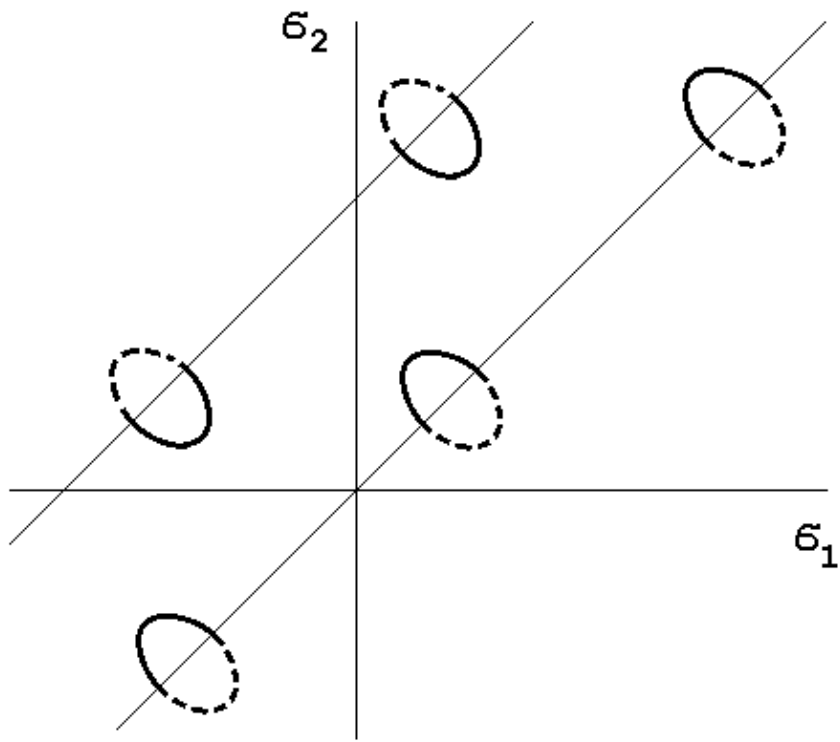
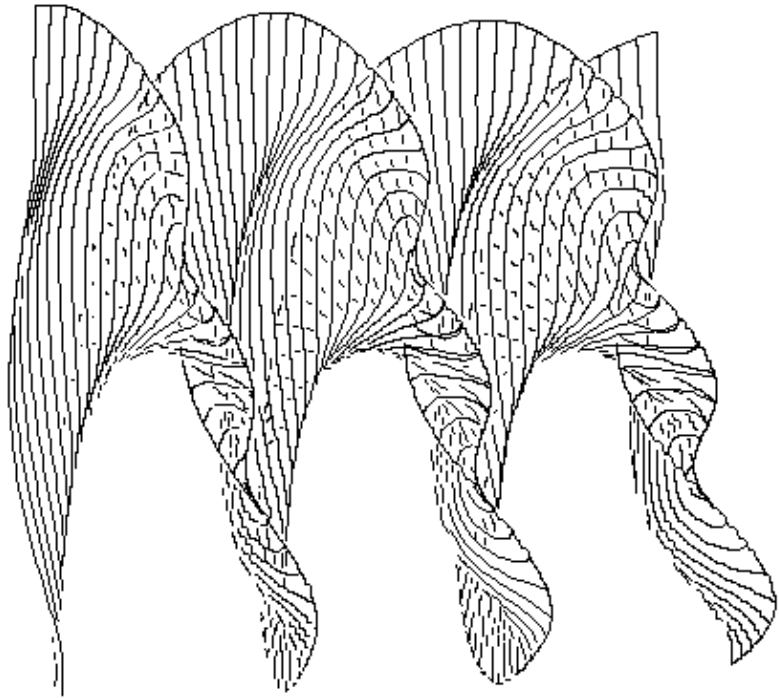
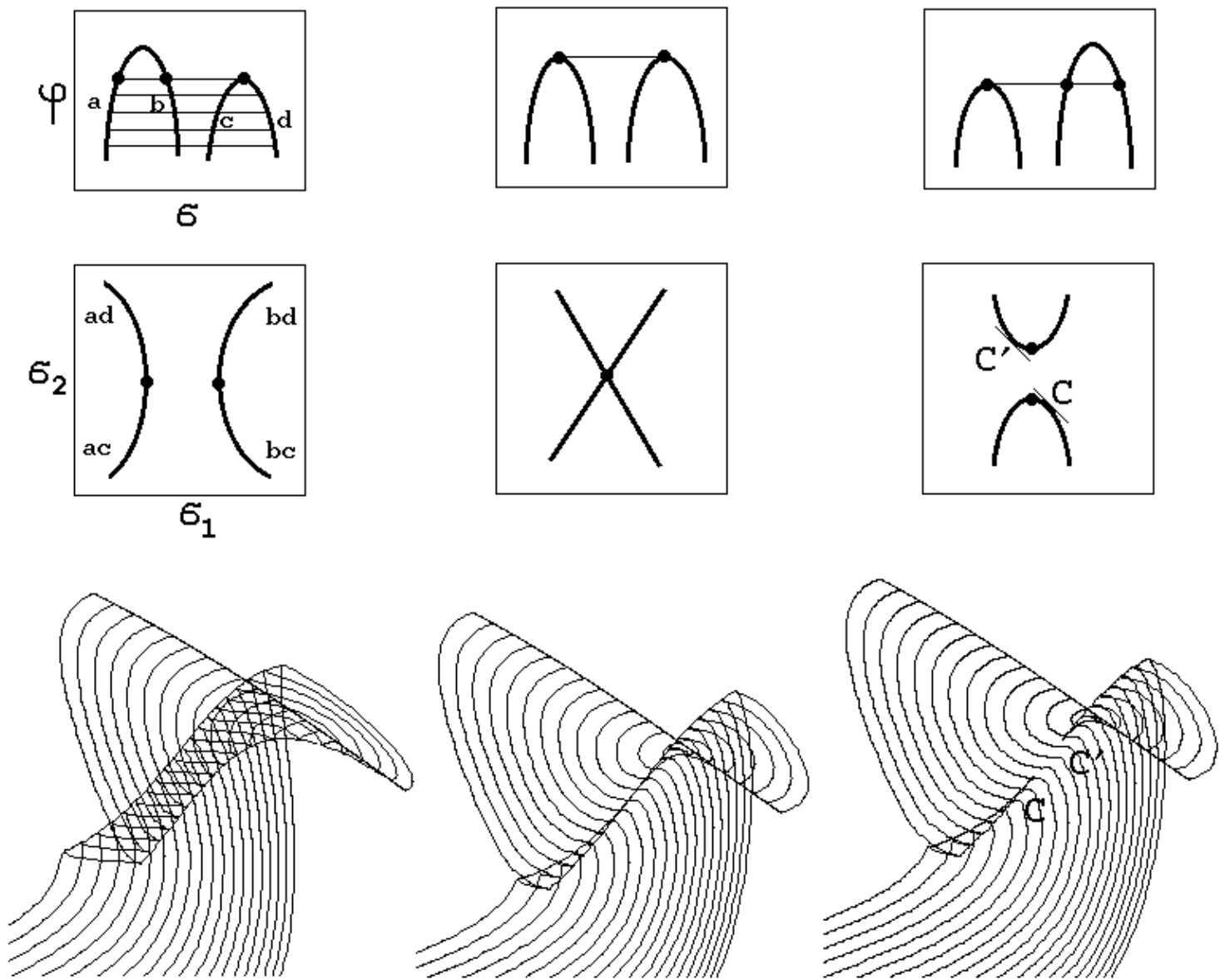
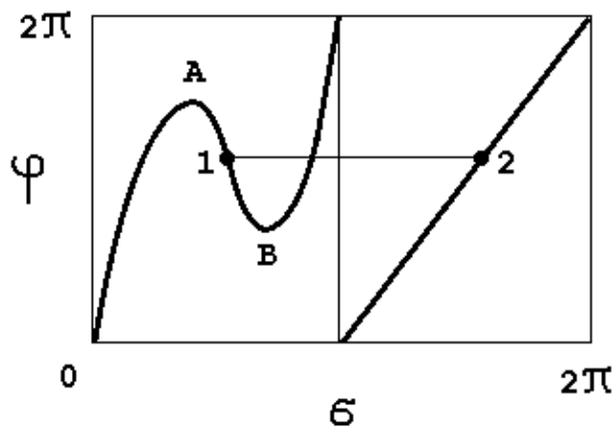
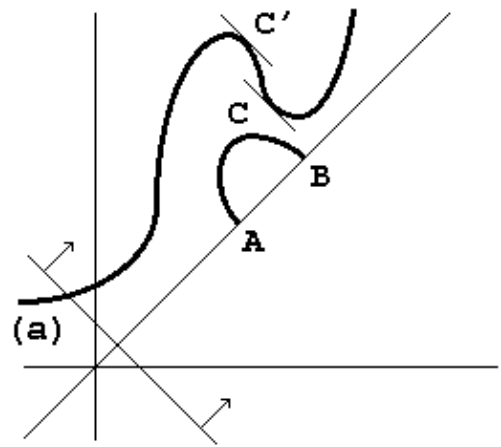
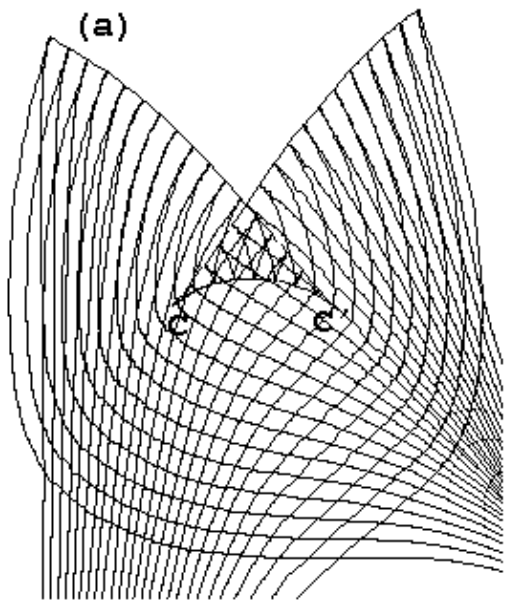
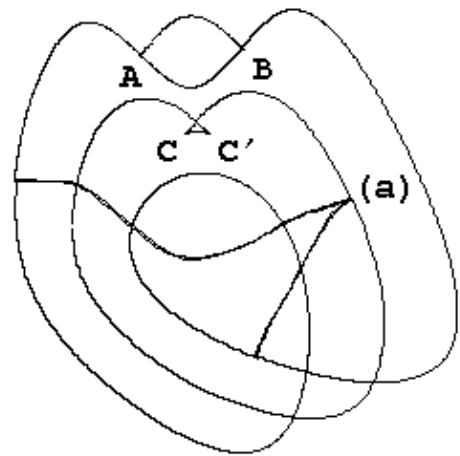
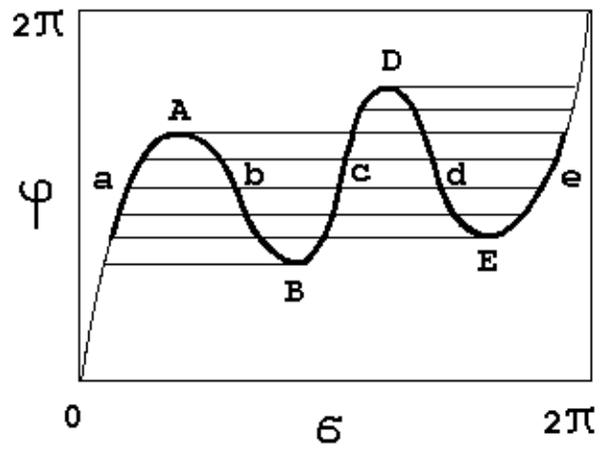
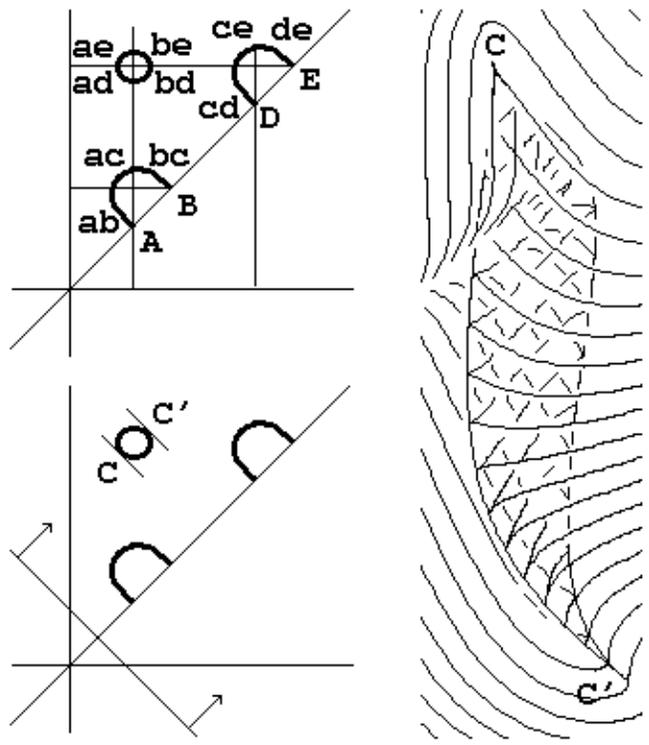
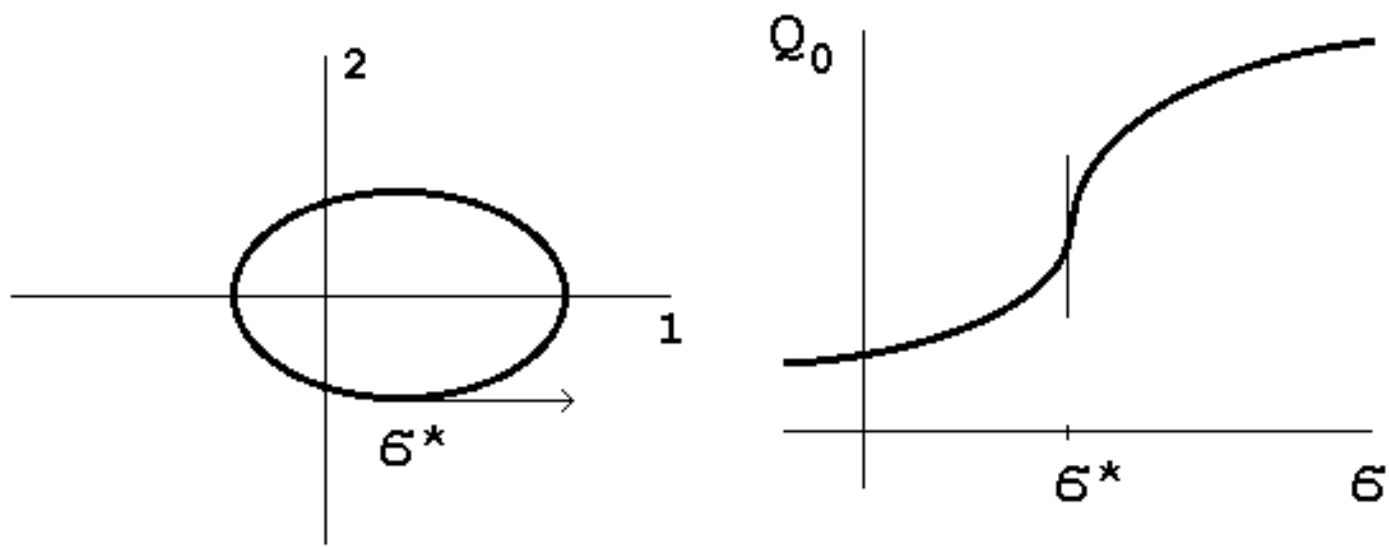
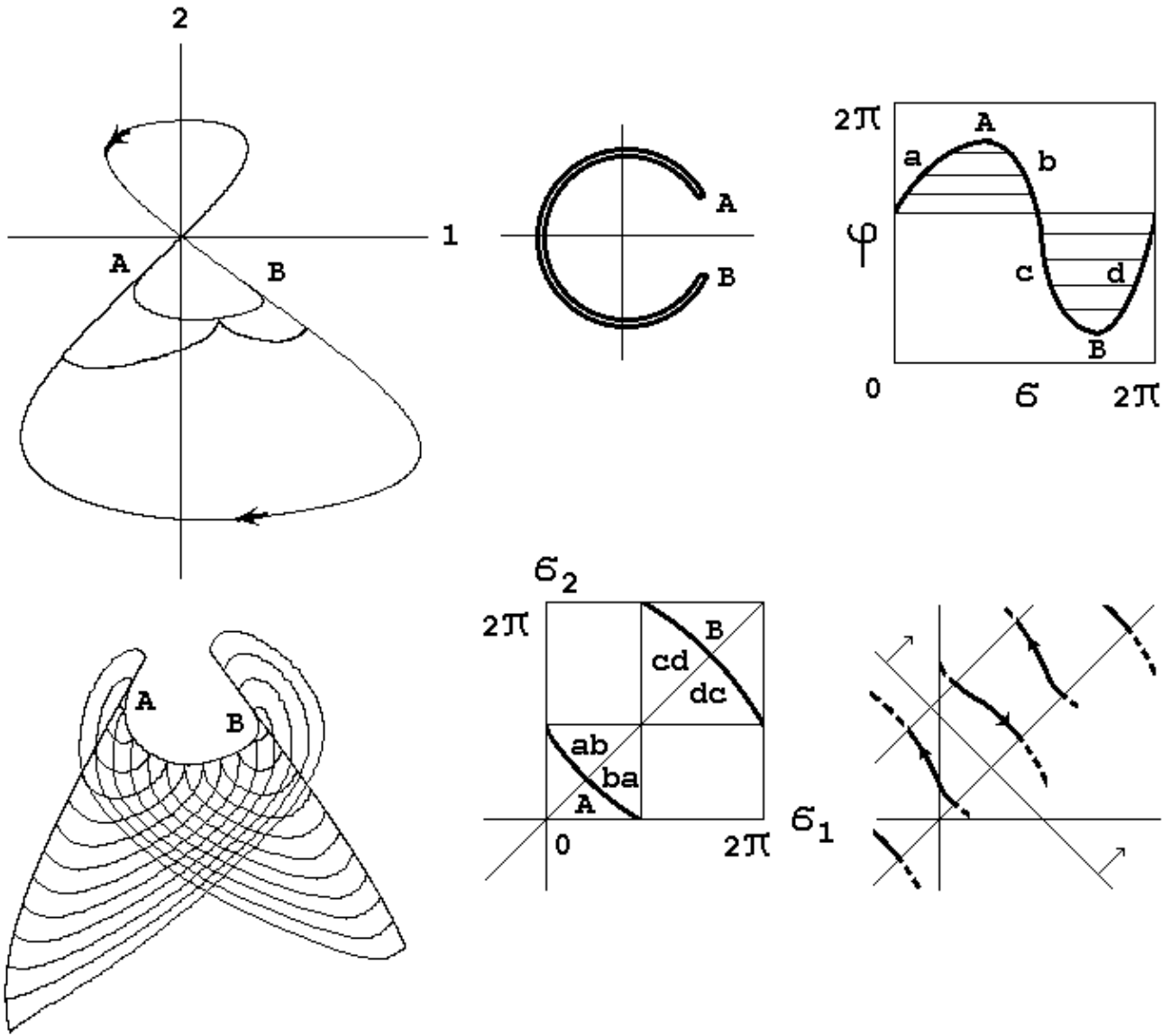
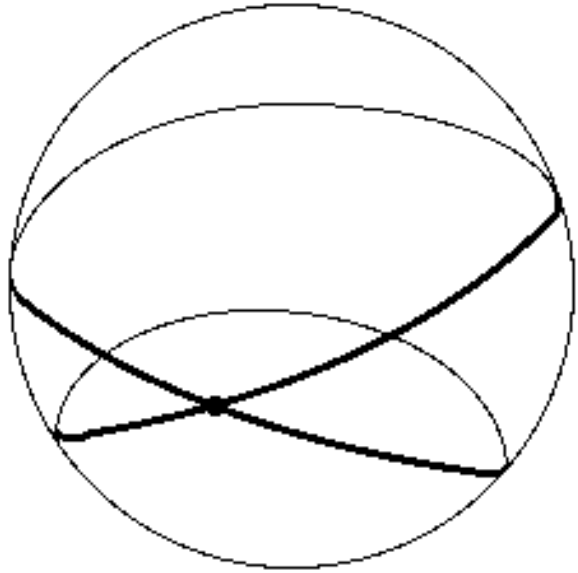
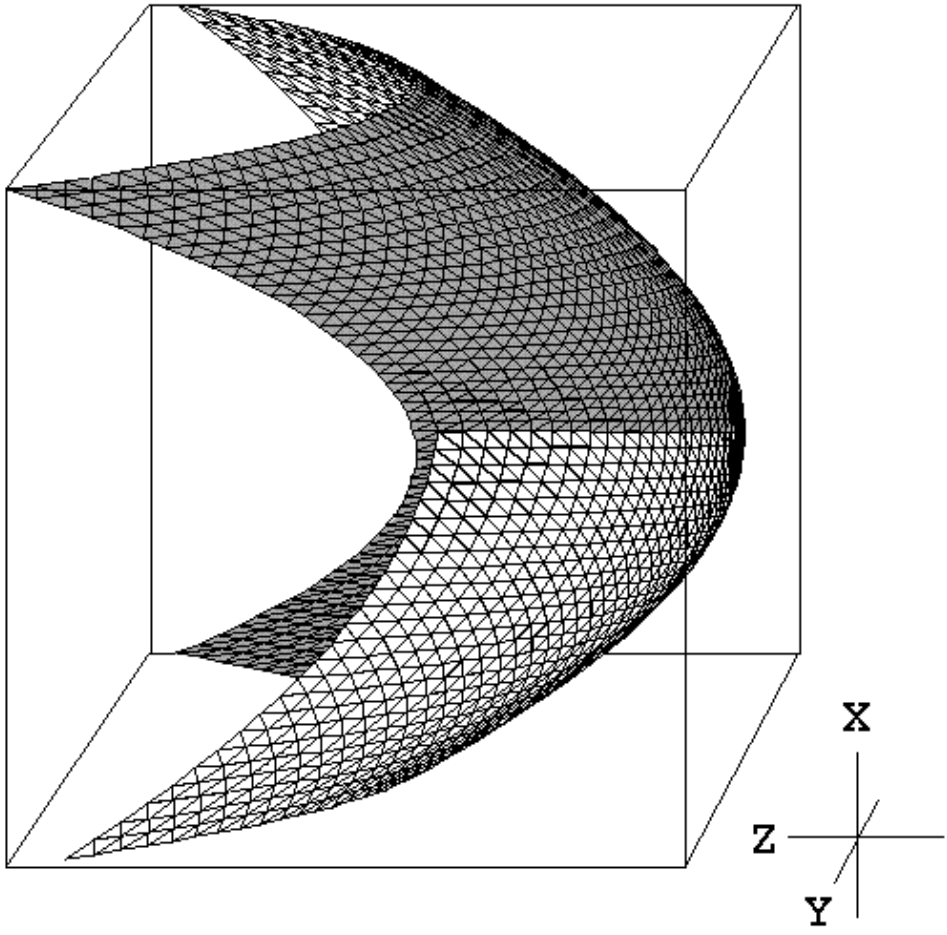
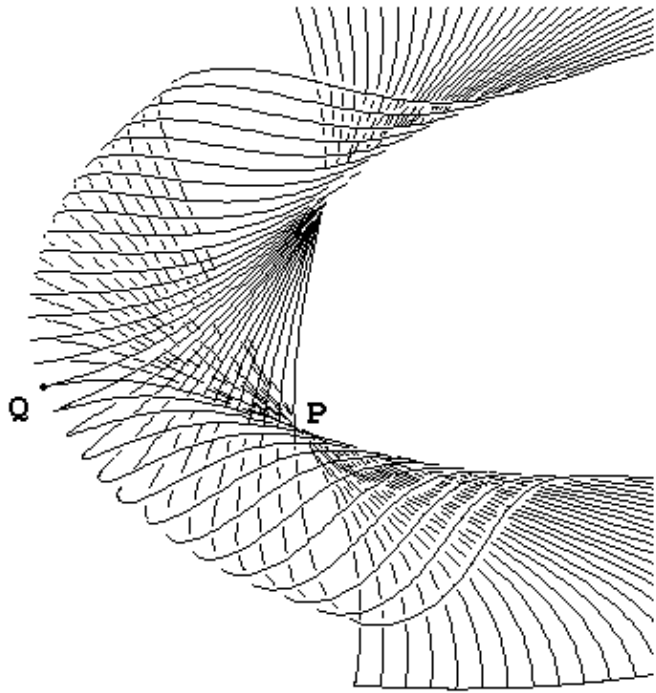
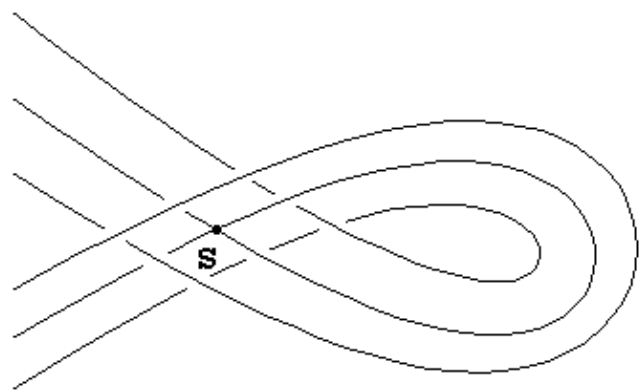
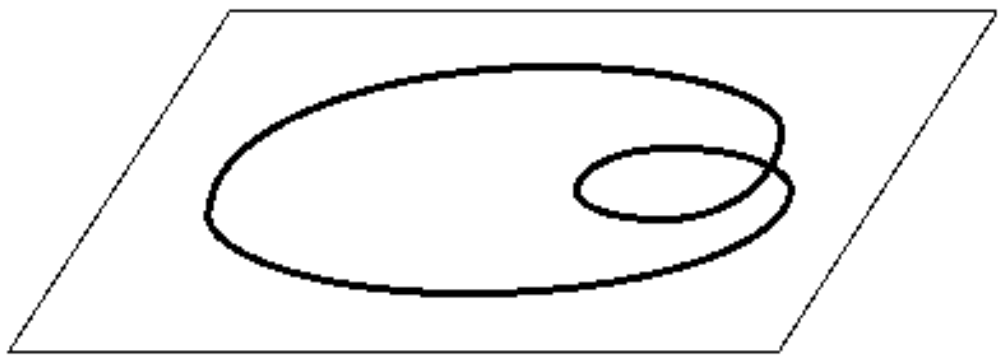
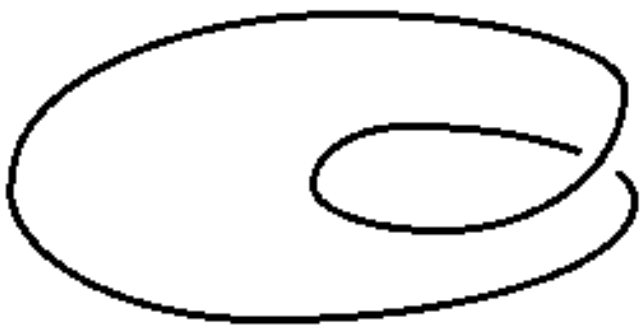
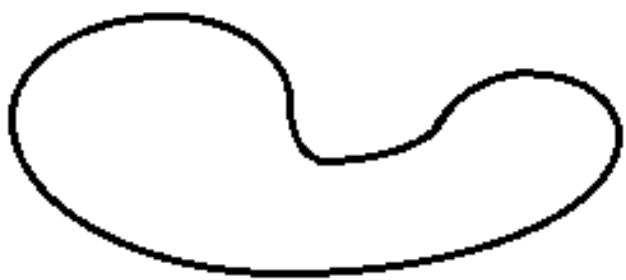
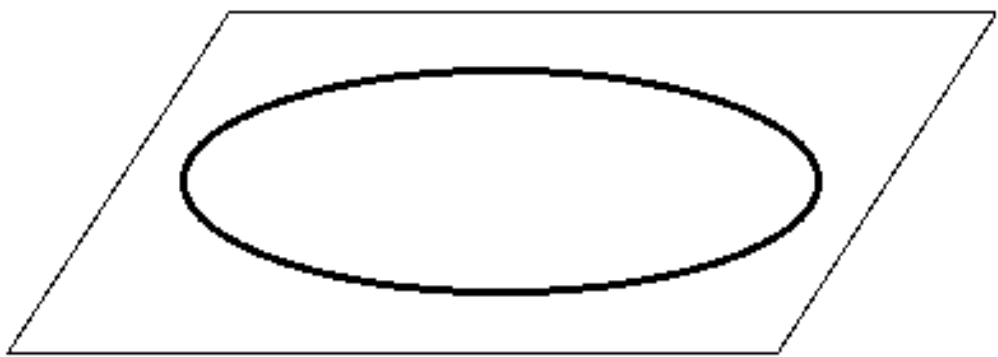
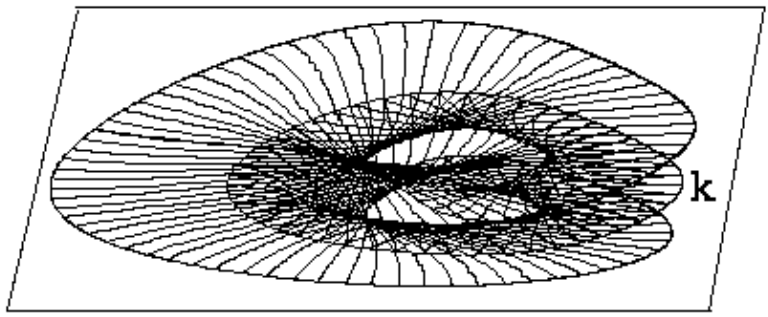
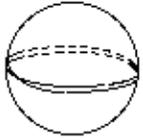
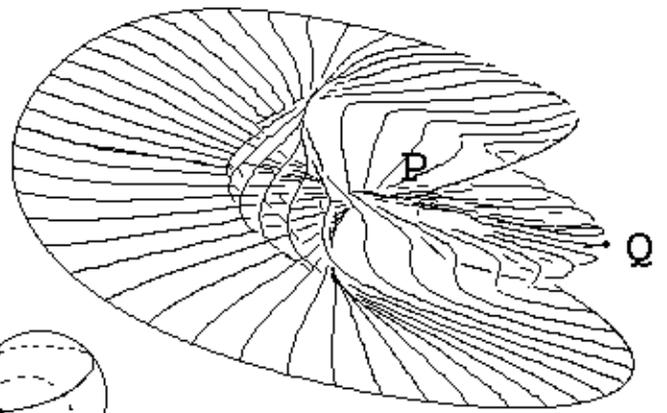
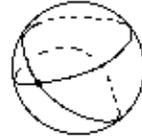
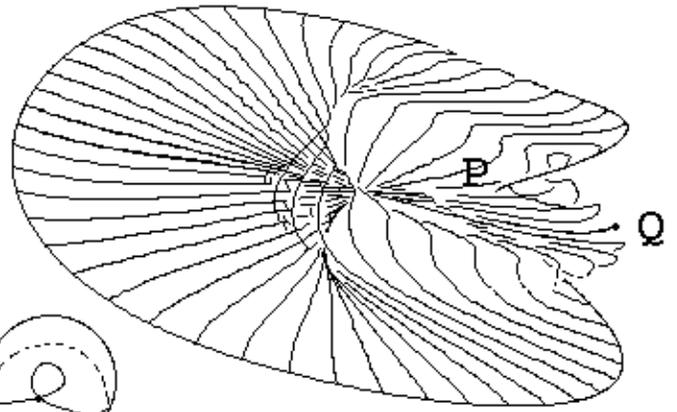
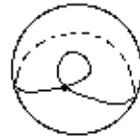
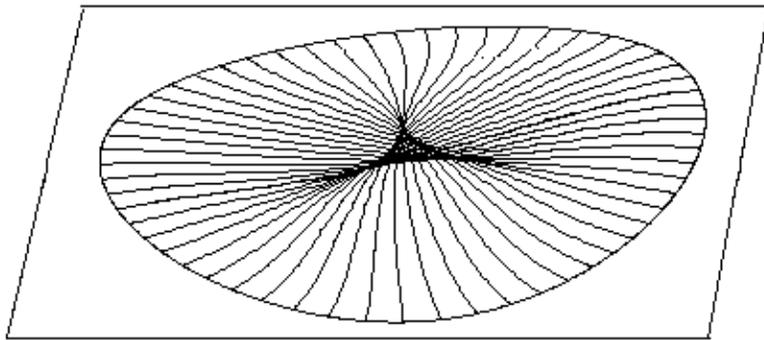
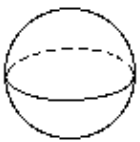
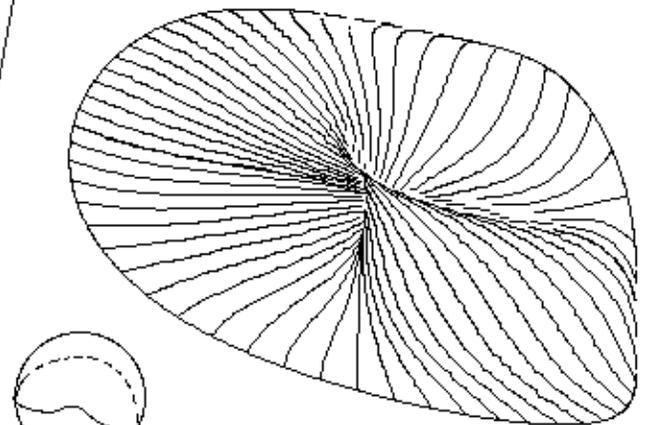
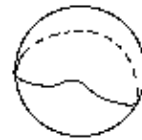
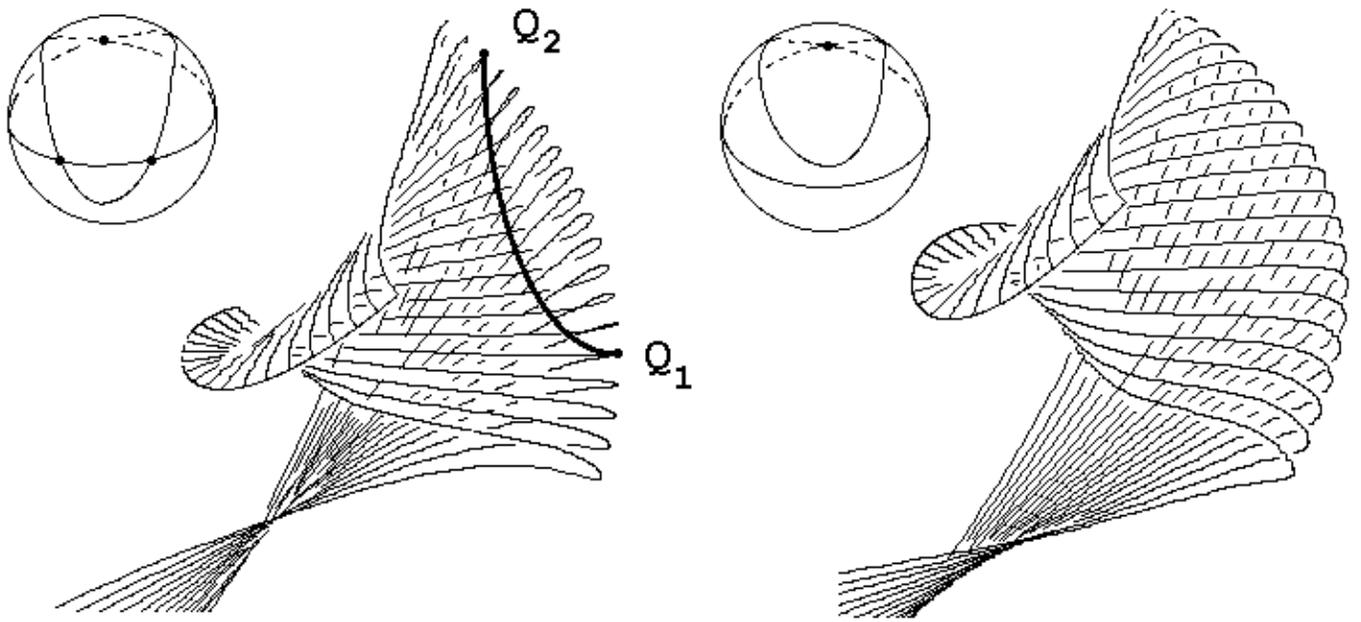
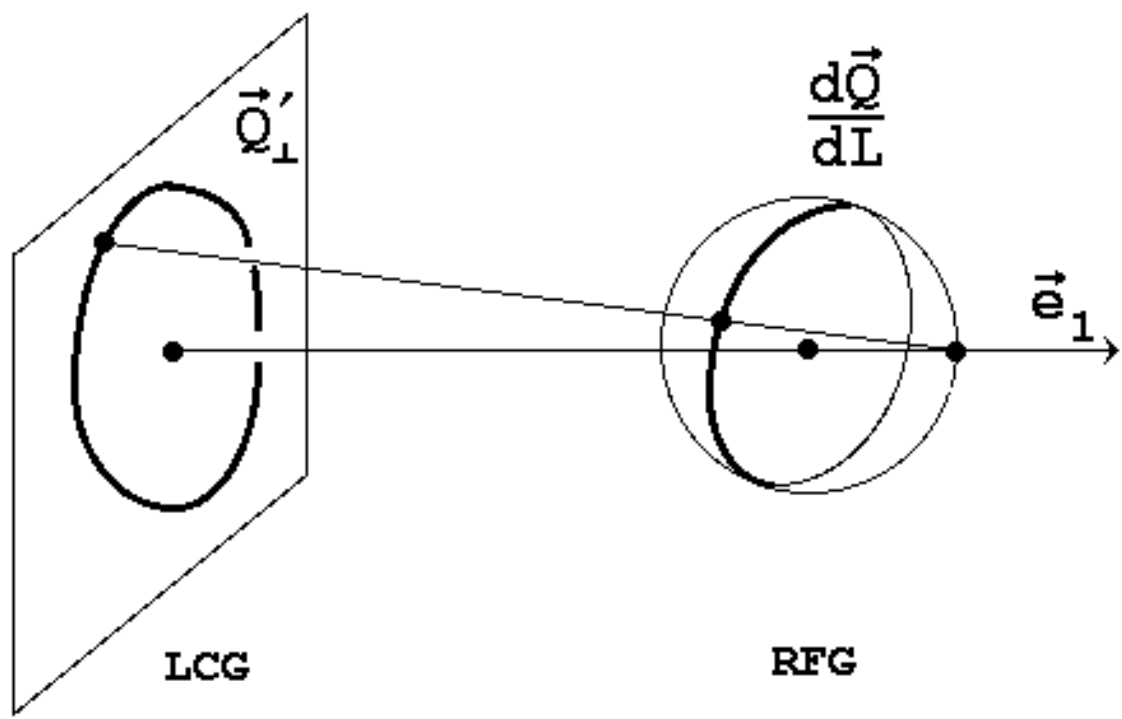
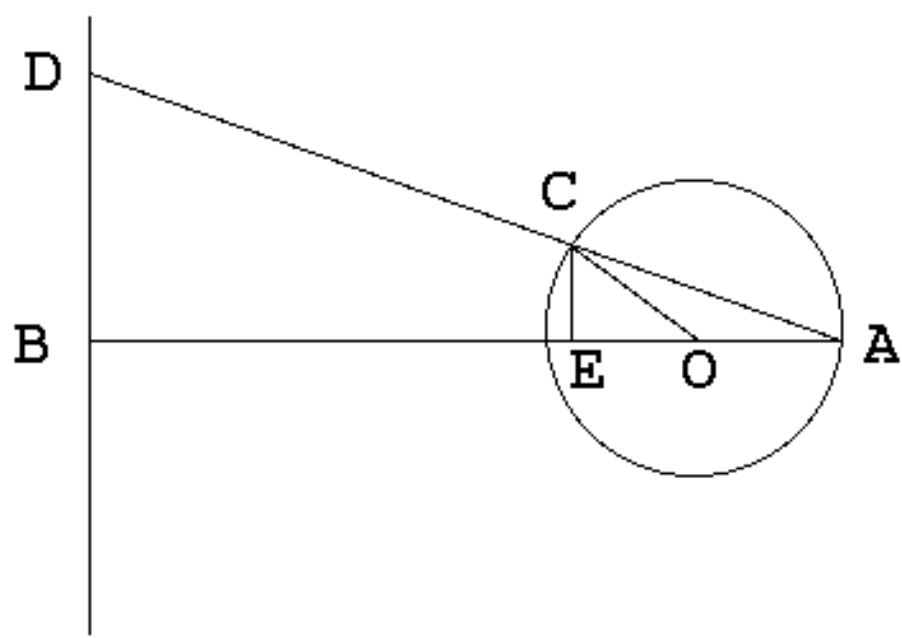
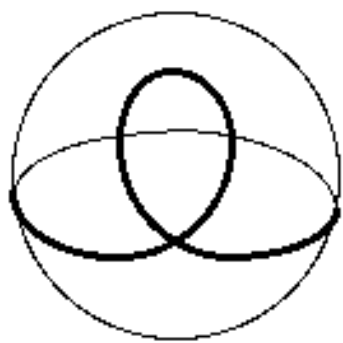
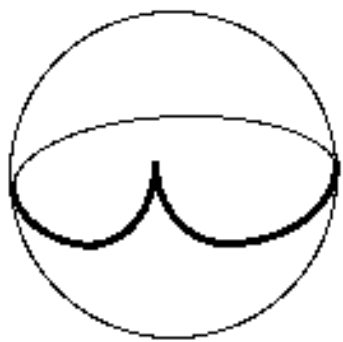
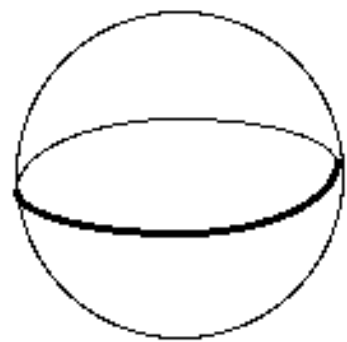
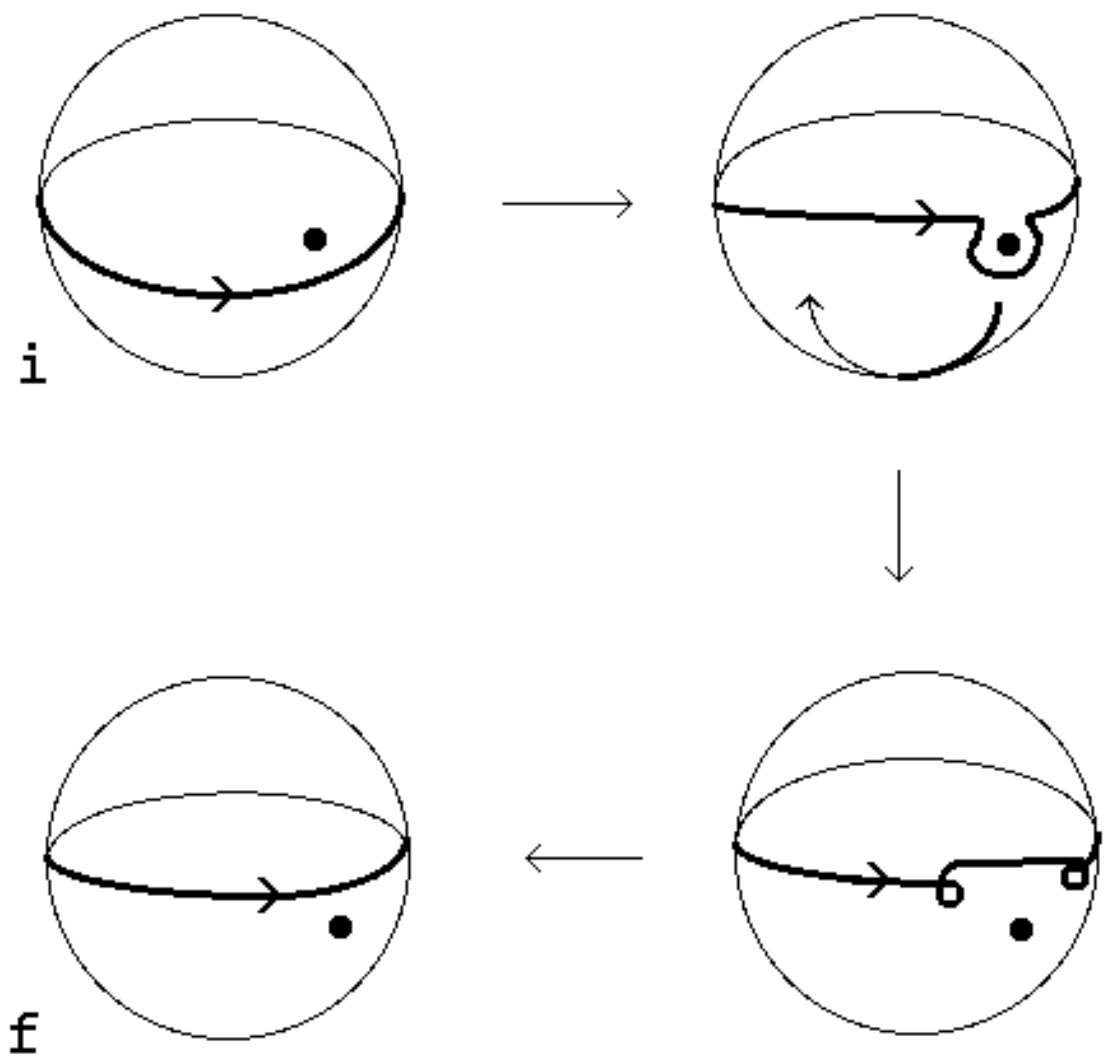
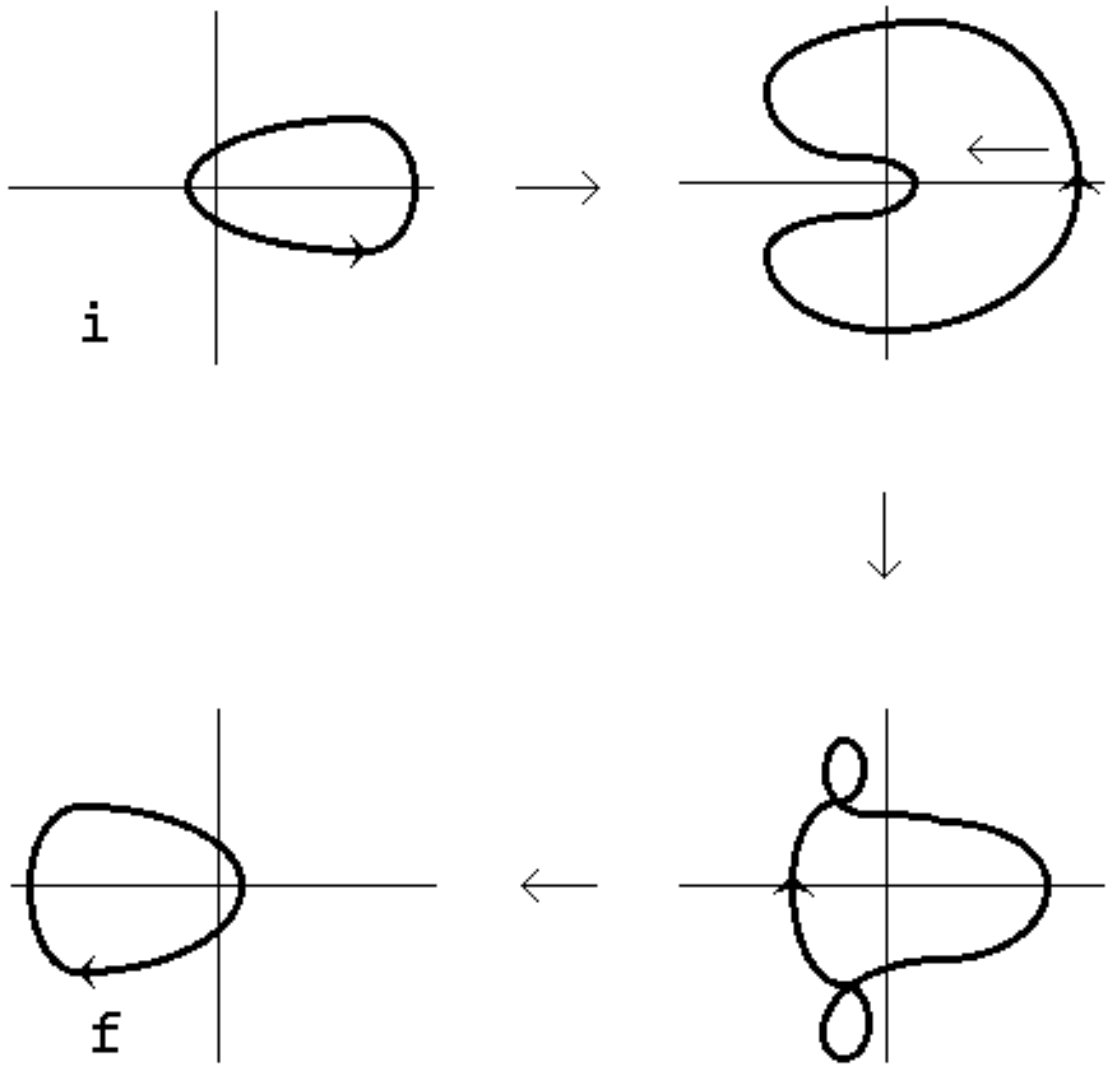
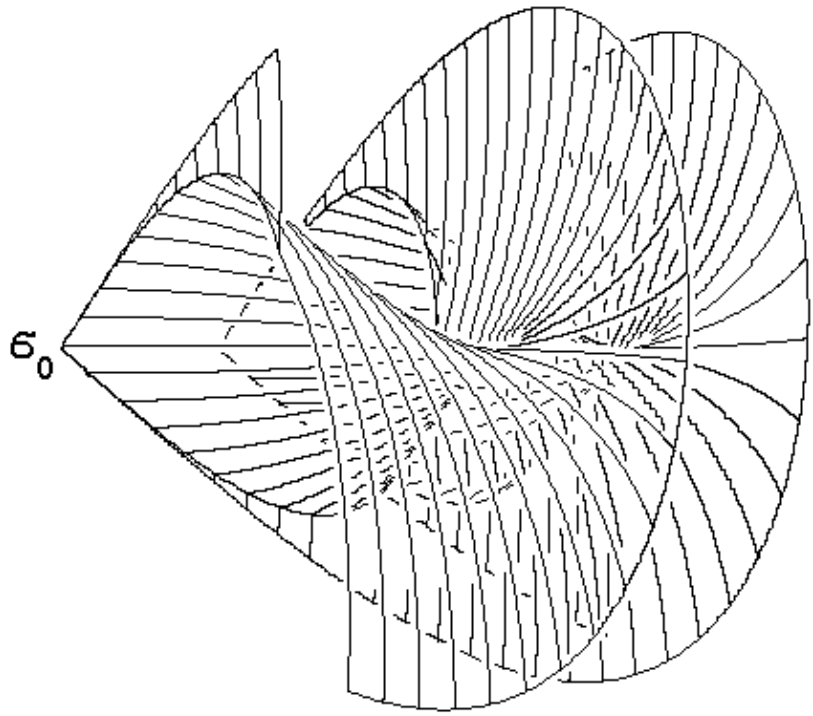
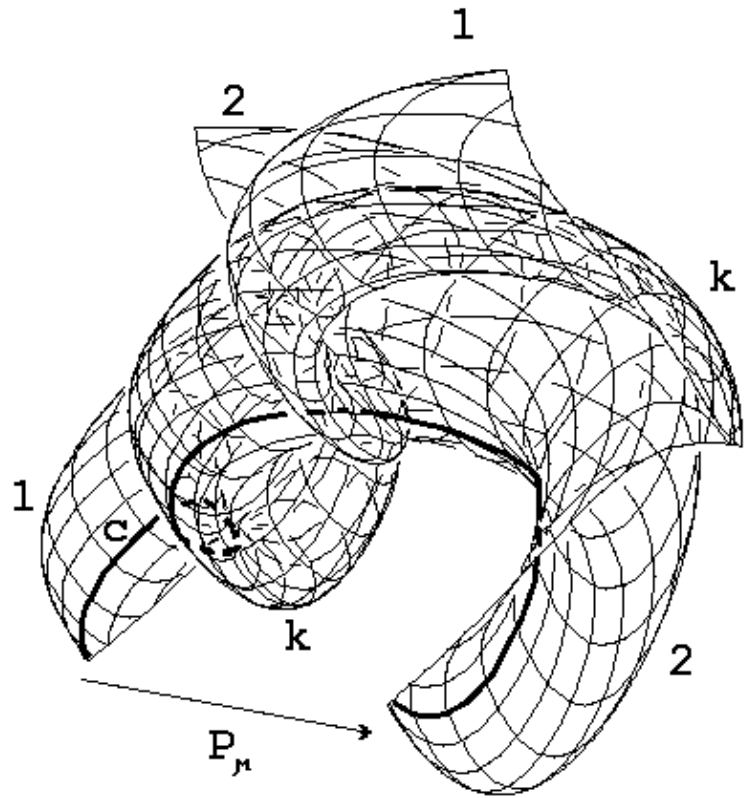
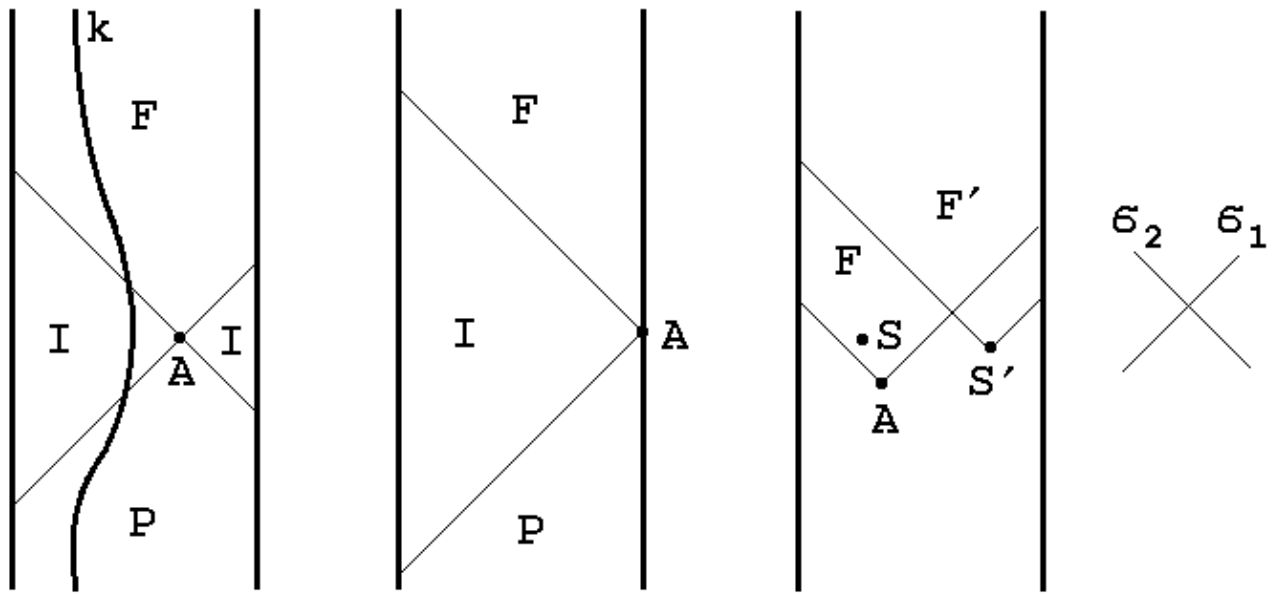
\end{document}